\documentclass[twocolumn,preprintnumbers,superscriptaddress,amsmath,amssymb,aps,prl,longbibliography]{revtex4-1}
\usepackage{braket}
\usepackage{miller}
\PassOptionsToPackage{textsize=tiny}{todonotes} % Sets todonotes textsize option to tiny when loaded by changes
\usepackage{changes}
\usepackage[colorlinks=true,allcolors=blue]{hyperref}

\usepackage[textsize=tiny]{todonotes}
\usepackage{graphics}% Include figure files

\usepackage[utf8]{inputenc}
\usepackage{multirow}
\usepackage{float}
\usepackage{bm}% bold math
\usepackage{bbold}
\usepackage{verbatim}
\usepackage{xfrac}
\usepackage{array}
\usepackage{siunitx}
\usepackage{ulem}
\usepackage{mathtools}
\usepackage{physics}

\newcommand{\pidx}[1]{{\mbox{\tiny $(#1)$}}}

\begin{document}
	
	\title{Enhanced Cavity Optomechanics with Quantum-well Exciton Polaritons}
	
	\author{N. Carlon Zambon}
    \email[Corresponding author.\\ Current affiliation: Photonics Laboratory, ETH Z\"{u}rich, CH-8093 Z\"{u}rich, Switzerland.\\]{carlonn@ethz.ch}
	\thanks{equal contribution.}
	\affiliation{Centre de Nanosciences et de Nanotechnologies, CNRS-Université Paris-Saclay, 91120 Palaiseau, France}
	% \altaffiliation{Current affiliation: Photonics Laboratory, ETH Z\"{u}rich, CH-8093 Z\"{u}rich, Switzerland}
	
	\author{Z. Denis}
	 \email[Corresponding author.\\ Current affiliation: Computational Quantum Science Laboratory, EPFL, CH-1015 Lausanne, Switzerland.\\]{zakari.denis@epfl.ch}
    \thanks{equal contribution.}
	\affiliation{Universit\'{e} Paris Cit\'{e}, CNRS, Mat\'{e}riaux et Ph\'{e}nom\`{e}nes Quantiques, F-75013 Paris, France}
	
	\author{R. De Oliveira}
	\affiliation{Universit\'{e} Paris Cit\'{e}, CNRS, Mat\'{e}riaux et Ph\'{e}nom\`{e}nes Quantiques, F-75013 Paris, France}
	
	\author{S. Ravets}
	\affiliation{Centre de Nanosciences et de Nanotechnologies, CNRS-Université Paris-Saclay, 91120 Palaiseau, France}
	
	\author{C. Ciuti}
	\affiliation{Universit\'{e} Paris Cit\'{e}, CNRS, Mat\'{e}riaux et Ph\'{e}nom\`{e}nes Quantiques, F-75013 Paris, France}
	
	\author{I. Favero}
	\affiliation{Universit\'{e} Paris Cit\'{e}, CNRS, Mat\'{e}riaux et Ph\'{e}nom\`{e}nes Quantiques, F-75013 Paris, France}
	
	\author{J. Bloch}
	\affiliation{Centre de Nanosciences et de Nanotechnologies, CNRS-Université Paris-Saclay, 91120 Palaiseau, France}

	\begin{abstract}
		Semiconductor microresonators embedding quantum wells can host tightly confined and mutually interacting excitonic, optical and mechanical modes at once. We theoretically investigate the case where the system operates in the strong exciton-photon coupling regime, while the optical and excitonic resonances are parametrically modulated by the interaction with a mechanical mode. Owing to the large exciton-phonon coupling at play in semiconductors, we predict an enhancement of polariton-phonon interactions by two orders of magnitude with respect to mere optomechanical coupling: a near-unity single-polariton quantum cooperativity is within reach for current semiconductor resonator platforms. We further analyze how polariton nonlinearities affect dynamical back-action, modifying the capability to cool or amplify the mechanical motion.
	\end{abstract}
	
	\maketitle
	
	%\setcounter{topnumber}{2}
	%\setcounter{bottomnumber}{2}
	%\setcounter{totalnumber}{4}
	%\renewcommand{\topfraction}{0.85}
	%\renewcommand{\bottomfraction}{0.85}
	%\renewcommand{\textfraction}{0.15}
	%\renewcommand{\floatpagefraction}{0.7}	
	
	%%%%%%%%%%%%%%%%%%%%%%%%%%%%%%%%	
	%
	%
	Optomechanical interactions represent an essential resource for augmented sensing techniques~\cite{Mason2019,Rossi2017,Halg2021}, in nonlinear optics~\cite{Dong2012,Purdy2013a,Chen2021,Hu2021}, and to investigate quantum phenomena in macroscopic systems~\cite{Purdy2013,Marinkovic2018,Delic2020,Ma2021}. Furthermore, coherent phonon scattering is an appealing route to implement microwave-to-optical transducers~\cite{Higginbotham2018,Mirhosseini2020,Arnold2020}, necessary to interface distant superconducting quantum hardware~\cite{Kimble2008,Barends2014,Ofek2016,Clerk2020}. To these ends, a key figure of merit is the single-photon quantum cooperativity $C_q=C_0/n_{\mathrm{th}}$, where $C_0$ is the single-photon cooperativity and $n_{\mathrm{th}}$ is the mechanical mode thermal occupation. It gauges the ability to coherently control the mechanical state with a single intracavity photon before environment-induced dephasing sets in~\cite{Aspelmeyer2014}. Maximizing $C_0=4g_0^2/\kappa\Gamma$ requires small optical ($\kappa$) and mechanical ($\Gamma$) dissipation rates and large single-photon optomechanical couplings ($g_0$), while $n_\mathrm{th}$ can be reduced using high-frequency mechanical resonators~\cite{Ding2011,Anguiano2017,Ren2020} and operating them at cryogenic temperatures~\cite{OConnell2010,Safavi-Naeini2012}. Recent works achieved a large cooperativity by engineering resonators with ultra-low mechanical and optical losses~\cite{Rossi2018,Ren2020}. A complementary approach is to devise solutions to enhance optomechanical interactions while working with modest optical and mechanical quality factors. Less stringent bandwidth limitations in optomechanical conversion are thereby imposed~\cite{Wang2012}, while suppressing optical heating and added noise~\cite{Ren2020}.
	%
	%Recent works achieved a large cooperativity by engineering resonators with ultra-low optical and mechanical  
	
	%
	In direct-bandgap semiconductors, photoelastic effects typically dominate optomechanical coupling~\cite{Baker2014} and are greatly enhanced near electronic resonances of the material~\cite{Feldman1968}. Moreover, in micromechanical resonators hosting quantum wells (QWs), electronic transitions can be tailored to boost carrier-mediated mechanical effects~\cite{Barg2018}. In both cases, an increase of optical absorption affects the cavity finesse and favors photothermal effects, while mechanical dephasing can be activated through photo-generated carriers~\cite{Lifshitz2000,Hamoumi2018}. In this context, GaAs-based resonators engineered to simultaneously confine photons, phonons and QW excitons offer an intriguing opportunity~\cite{Fainstein2013,Rozas2014,Villafane2018}: in the strong exciton-photon coupling regime the system hosts hybrid quasi-particles, or polaritons, that share properties of both of their constituents~\cite{Carusotto2013}. Polariton modes are spectrally separated from the exciton-induced absorption peak, enabling large optical quality factors, while their excitonic component is extremely sensitive to strain fields owing to the large GaAs deformation potential~\cite{Bardeen1950,Bir-Pikus1974,Piermarocchi1996}, thus prospecting strong optomechanical interactions. Polaritons are bosonic quasi-particles which can form non-equilibrium condensates~\cite{Kasprzak2006,Deng2010}, while strong exciton-mediated nonlinearities enable the occurrence of superfluid behaviours~\cite{Amo2009,Lerario2017}, dissipative phase transitions~\cite{Rodriguez2017,Fink2018} and parametric processes~\cite{Kuznetsov2020,CarlonZambon2020}. Optomechanical interactions offer an additional degree of freedom for quantum fluids of light foreshadowing new possibilities. Recent experiments showing mechanical lasing driven by a polariton condensate~\cite{Chafatinos2020}, the electrical actuation of polariton-phonon interactions~\cite{Kuznetsov2021}, and giant polariton-induced bulk photoelastic effects~\cite{Jusserand2015,Kobecki2021}, support this intuition, and call for the development of an unifying theoretical framework. Early works that established the foundations of polariton optomechanics, either focused on static effects and neglected the role of exciton-phonon and exciton-exciton interactions~\cite{Kyriienko2014}, or studied a two-level atom strongly coupled to an optomechanical cavity~\cite{Restrepo2014,Restrepo2017}.
	
	Here we model the tripartite interaction of light, QW excitons, and sound in semiconductor microresonators. In the strong exciton-photon coupling regime, we show that such interaction generates a radiation-pressure type Hamiltonian, with photons replaced by polaritons, and with an effective optomechanical coupling given by the weighted sum of the photon-phonon and exciton-phonon couplings. We provide analytical derivations of the effective optomechanical coupling for three resonator architectures: when considering parameters complying with current GaAs technologies, because of the giant exciton-phonon contribution, we show that a near-unity cooperativity can be obtained for a single polariton excitation. Finally, we investigate how polariton nonlinearities modify dynamical back-action via squeezing. 
	
	%%%%%%%%%%%%%%%%%%%%%%%%%%%%%%	
	
\begin{figure}[t]
		\centering
		\includegraphics[trim=0cm 0cm 0cm 0cm, width=86mm]{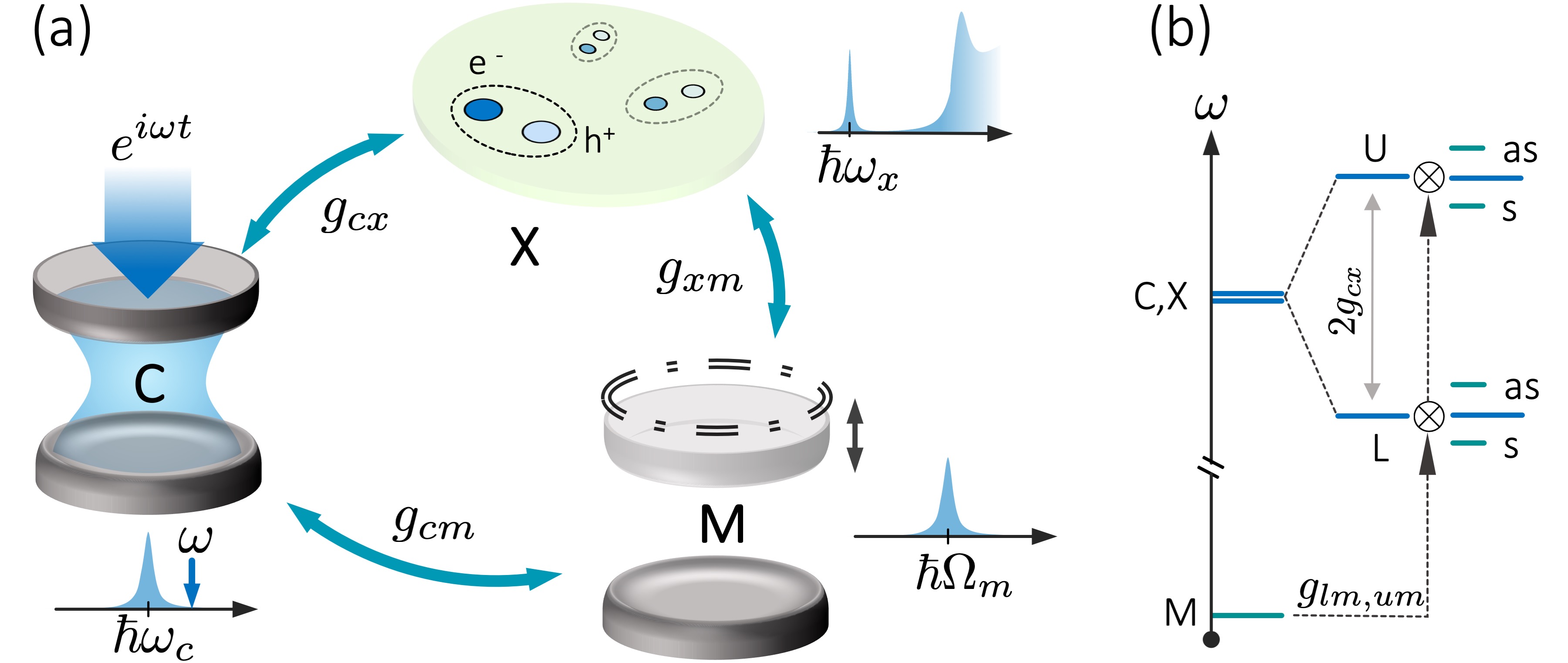}
		\caption{(a) A resonator supports optical and mechanical modes with respective frequencies $\omega_c$ and $\Omega_m$. It also embeds a quantum well (QW) with excitonic resonance at $\omega_x\approx\omega_c$. The cavity is driven by a laser with frequency $\omega$. We denote $g_{ij}$ the pairwise couplings among the modes. (b) Energy level diagram. Strong exciton-photon coupling results in a Rabi splitting $\Omega_R\approx 2g_{cx}$ between the upper (U) and lower (L) polariton normal modes. The mechanical mode modulates the polariton resonances through the effective couplings $g_{lm}$ and $g_{um}$, producing Stokes (s) and anti-Stokes (as) sidebands.}
		\label{fig:TheSystem}
	\end{figure}

	\textit{Model}~---~We consider the coupled dynamics of three bosonic fields describing the optical cavity mode, QW excitons, and a mechanical degree of freedom. We restrict ourselves to a single-mode scenario, leaving the generalization to the multi-mode case to forthcoming works. Notice that the exciton bosonization implies that we neglect electronic phase-space filling effects~\cite{Carusotto2013}. The large exciton effective mass in GaAs enables us neglecting its dispersion for all in-plane optical wavevectors \cite{Panzarini1999}. The bare system Hamiltonian reads ($\hbar=1$)
	\begin{equation}
		\hat{H}_0=\omega_{c} \hat{a}^{\dag}\hat{a} + \omega_{x} \hat{d}^{\dag}\hat{d} + \frac{g_{xx}}{2}\hat{d}^{\dag}\hat{d}^{\dag}\hat{d}\hat{d} + \Omega_m\hat{b}^{\dag}\hat{b},
	\end{equation}  
	where $\omega_c$, $\omega_x$ and $\Omega_m$ denote the cavity (C), exciton (X) and mechanical (M) resonance frequencies, associated to the bosonic ladder operators $\hat{a}$, $\hat{d}$ and $\hat{b}$, while the anharmonic term proportional to $g_{xx}$ takes into account exciton exchange interactions~\cite{Ciuti1998}. The couplings among the three modes are captured by 
	\begin{equation}
		\hat{H}_I=g_{cx} (\hat{a}^{\dag}\hat{d} + \hat{d}^{\dag}\hat{a})-g_{cm}\hat{a}^{\dag}\hat{a}(\hat{b}+\hat{b}^{\dag}) -  g_{xm}\hat{d}^{\dag}\hat{d}(\hat{b}+\hat{b}^{\dag}).
	\end{equation}
	The first term describes dipole photon-exciton interactions ($\omega_c\approx\omega_x \gg g_{cx}$), while the other two describe the parametric modulation of the C and X resonances actuated by the mechanical field ($g_{cm,xm}\ll\omega_{c,x}$)~\cite{Aspelmeyer2014}. The coupling $g_{cm}$ contains both geometric-deformation and photo-elastic effects~\cite{Baker2014}, while $g_{xm}$ accounts for the exciton-phonon interaction via the deformation potential~\cite{Bir-Pikus1974}, see Fig.~\ref{fig:TheSystem}\,(a). We denote the C, (nonradiative) X and M decay rates $\kappa_c$, $\kappa_x$ and $\Gamma$.
	In the strong C-X coupling regime ($g_{cx}\gg\kappa_{c,x}$), the normal modes of the light-matter Hamiltonian in the single-excitation subspace form the relevant basis. The bare X and C modes hybridize yielding the lower (L) and upper (U) polariton resonances $2\omega_{l,u}=(\omega_c+\omega_x) \mp \sqrt{\delta^2_{cx}+4g^2_{cx}}$, with $\delta_{cx}=(\omega_c-\omega_x)$.  Polaritons are described by ladder operators $(\hat{u},\hat{l}\,)^T=\mathcal{R}[\theta_{cx}](\hat{d},\hat{a})^T$ where $\mathcal{R}[\theta_{cx}]$ is a rotation with mixing angle $\theta_{cx}$ satisfying $\cos2\theta_{cx}=-\delta_{cx}/\sqrt{\delta^2_{cx}+4 g^2_{cx}}$. As a result, phonons effectively couple to L via $g_{lm}=(g_{xm} \sin^2\theta_{cx} + g_{cm} \cos^2\theta_{cx})$, and to U via $g_{um}=(g_{xm}\cos^2\theta_{cx} + g_{cm} \sin^2\theta_{cx})$. In the polariton basis $\hat{H}_0+\hat{H}_I = \hat{H}_l+\hat{H}_u+\hat{H}_{m}+\hat{H}_{lu}$, where 
	\begin{equation}\label{eq:EffectiveLPUPHam}
		\hat{H}_{j=(l,u)}=\left[\omega_{j} + \frac{\chi_{j}}{2}(\hat{n}_{j}-1) - g_{jm}(\hat{b}+\hat{b}^{\dag}) \right]\hat{n}_{j}
	\end{equation}
	describes interacting polaritons in the L and U branches that are parametrically coupled to a mechanical mode.  Here $\hat{n}_{j}$ denote number operators, $\chi_{l}= g_{xx} \sin^4\theta_{cx} $, $\chi_{u}=g_{xx} \cos^4\theta_{cx}$, $\hat{H}_{m}=\Omega_m\hat{b}^{\dag}\hat{b}$ and $\hat{H}_{lu}= - g_{lu} (\hat{b}+\hat{b}^{\dag})(\hat{l}^{\dag}\hat{u}+\hat{u}^{\dag}\hat{l})$ is a mechanically-assisted coupling between the L and U polariton branches, with $g_{lu} = \sin(2\theta_{cx})(g_{xm}-g_{cm})/2$. We sketch the energy levels for the coupled CXM system in Fig.~\ref{fig:TheSystem}\,(b). Interestingly, $\hat{H}_{lu}$ describes a coherent three wave-mixing  among the polariton branches mediated by phonons, becoming resonant as the mechanical frequency matches the normal mode splitting $\Omega_m=\sqrt{\delta^2_{cx}+4g^2_{cx}}$, enabling a coherent population transfer~\cite{Vyatkin2021}. Hereafter, we consider driving coherently C using a narrow-band laser of frequency $\omega$, see Fig.~\ref{fig:TheSystem}\,(a).

\begin{figure}[t]
		\centering
		\includegraphics[trim=0cm 0cm 0cm 0cm, width=86mm]{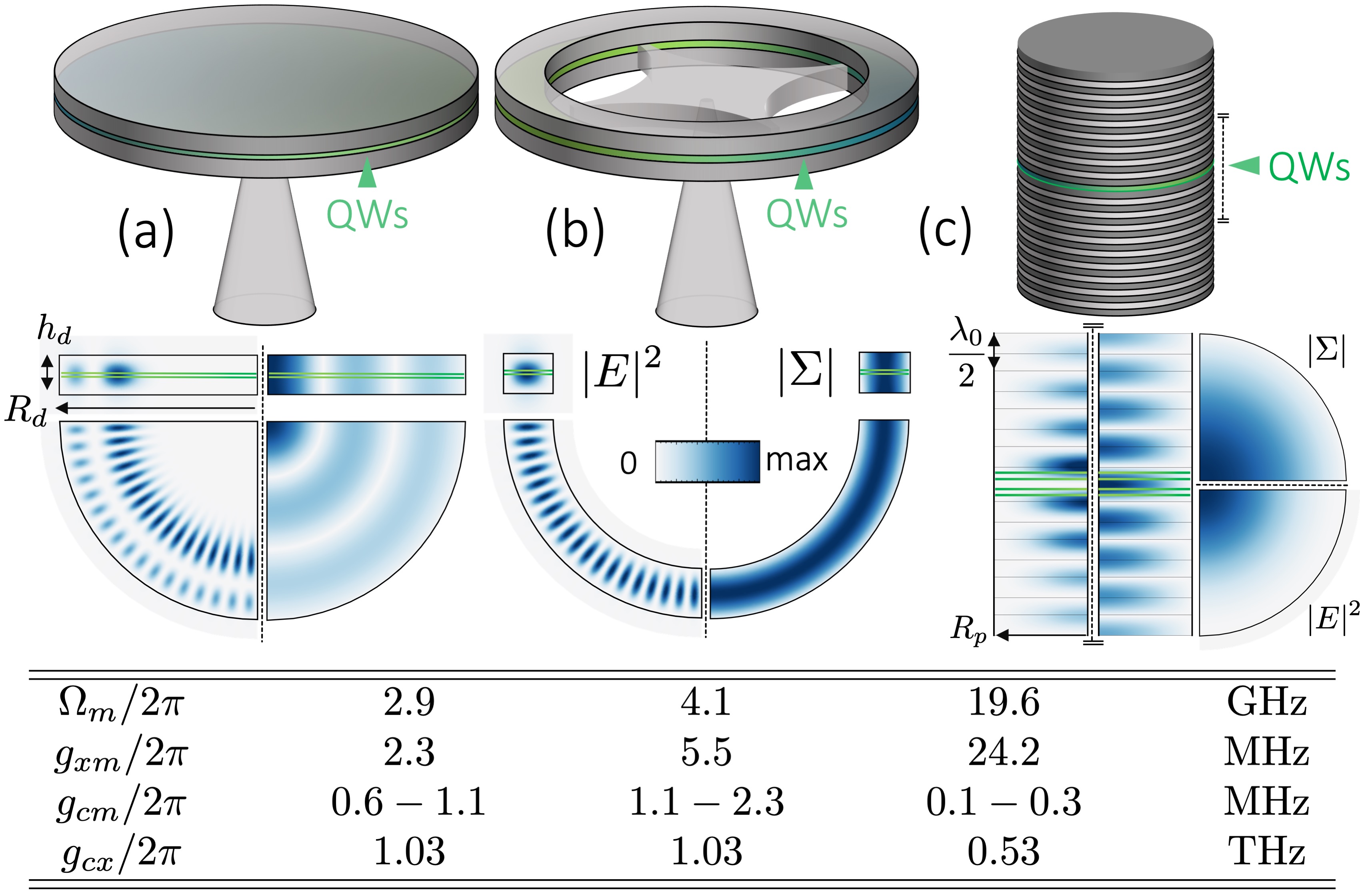}
		\caption{Sketch of a disk (a), ring (b) and pillar (c) microresonator. Here (a,b) are $\SI{0.2}{\um}$ thick with an outer radius $R_d = \SI{2}{\um}$; in (b) the inner radius is $\SI{1.5}{\um}$. (c) is $\SI{2.6}{\um}$ in diameter and defined by two GaAs/AlAs distributed Bragg reflectors (DBRs) and $\lambda/2$ GaAs spacer. All structures are adjusted to yield optical resonances near $E_x=\SI{1.463}{\eV}$ (for a $\SI{8}{\nm}$ thick $\mathrm{In_{0.05}Ga_{0.95}As}$ QW~\cite{Note1}). Normalized intensity profiles of a cavity mode ($|E|^2$) and strain field $(|\Sigma|)$ are shown for in-plane and orthogonal cuts. The QW positions are highlighted with solid lines: in (a,b) two QWs are displaced by $\pm\SI{15}{\nm}$ from the cavity field anti-node ($\eta_S\approx 1$), whereas in (c) $4$ QWs displaced by $\pm (15,39)\,\mathrm{nm}$ from the center of the spacer $\eta_{S}=(0.93,0.62)$. The calculated M frequency ($\Omega_m/2\pi$) and coupling rates $(g_{ij}/2\pi)$ are listed for the selected modes. We list $g_{cm}/2\pi$ for a cavity mode detuning of $40$--$\SI{10}{\nm}$ from the GaAs bandgap. Information on the optical and mechanical decay rates can be found in \cite{Note1}.}
		\label{fig:threeExamples}
	\end{figure}

	\textit{Electromechanical coupling}~---~Exciton-phonon coupling stems from the  strain-induced perturbation of the semiconductor band structure~\cite{Bir-Pikus1974}. The resulting exciton energy shift is given by $U(\vb*{r}_e,\vb{r}_h) = a_e \Sigma_n(\vb*{r}_e) - a_h \Sigma_n(\vb*{r}_h)$~\cite{Piermarocchi1996}, where $a_{e,h}$ are the electron and hole deformation potentials and $\Sigma_n(\vb*{r}) = \nabla \cdot \vb*{u}_n(\vb*{r})$ denotes the volumetric strain at the position $\vb*{r}$ imputable to a phonon in the mode $n$ associated to the displacement field $\vb*{u}_n(\vb*{r})$. Upon tracing $U(\vb*{r}_e,\vb*{r}_h)$ in the exciton and phonon basis, as the exciton Bohr radius is much smaller than the phonon wavelength in the QW plane (cf. \footnote{See \hyperref[app:A]{Supplemental Material}, including Refs.~\cite{Anselm1955,Herring1956,Altland2010,Paul1991,Zubkov2004,Bastard1992,Levinshtein1996,Adachi1982,Moore1990,Andreani1995,Strutt1878,Rayleigh1910,Rayleigh1914,Landau2008,Parrain2014,Hao2007,Anetsberger2008,Girlanda1981,Whittaker2018,Yeh1979,Marte1997,Hauer2013,Karl2009,Kippenberg2005,Drummond1980,Clark2017,Marquardt2007}, for details on the analytical derivation of the exciton-phonon coupling in the three microresonator architechtures, on practical limitations to the optical and mechanical quality factors, on shallow quantum well excitons and on the theory of polariton dynamical backaction.}), the electromechanical coupling reduces to
	\begin{equation}\label{eq:approxGxm}
		g_{xm}^{(n,m)} \approx(a_h - a_e)\int_{S}\mathrm{d}\vb*{R}\lvert E_{m}(\vb*{R})\rvert^2\Sigma_n(\vb*{R}, z_\mathrm{QW}),
	\end{equation}
	where $\vb*{R}$ is the vector spanning the QW plane over the horizontal cross-section $S$ of the resonator, $z_\mathrm{QW}$ is the position of the QW along the vertical axis and $E_m(\vb*{R})$ is the electric field distribution for the $m$th optical mode at the QW plane. In the strong coupling regime the optical mode enters the overlap integral as the exciton density is dictated by the cavity field profile \cite{Panzarini1999}. We now evaluate Eq.~\eqref{eq:approxGxm} for the three microresonator geometries presented in Fig.~\ref{fig:threeExamples}: disk (a), ring (b) and pillar (c) microresonators. For each architecture, Fig.~\ref{fig:threeExamples} shows representative profiles of the optical and strain fields presenting a near-optimal overlap. We could find analytical expressions for the mode envelopes and recast Eq.~\eqref{eq:approxGxm} as $\hbar g^{(n,m)}_{xm}=(a_e-a_h)(x_{\mathrm{ZPF}}k_m) \mathcal{I}_g \eta_{S}$ where $x_{\mathrm{ZPF}}$ is the zero-point fluctuation amplitude, $k_m$ is the phonon wave-vector, $\mathcal{I}_g$ is a geometric overlap integral and $\eta_{S}$ is the ratio between the peak value of the strain in the QW plane and its maximum (hence $\eta_S<1$). As $|a_e-a_h|\approx \SI{9.7}{\eV}$ for GaAs~\cite{Piermarocchi1996}, we expect $g_{xm}$ to be larger than $g_{cm}$, sharing a similar expression but a prefactor proportional to $\hbar\omega_c$~\cite{Baker2014,Anguiano2018}. 

	In \cite{Note1}, we compute $\mathcal{I}_g$ for any radial breathing mode (RBM) and any whispering gallery mode (WGM) for resonators (a,b), while for the fundamental optical and longitudinal breathing mode of (c) we find
	\begin{equation}\label{eq:GeomIntegralPill}
		\mathcal{I}_{g}  \approx \beta_0 \exp\left(\Delta n/2n_\mathrm{eff}\right)/\pi J_{1}(\alpha_{01}).
	\end{equation}
	Here $\beta_0\approx 1.18$, $J_n(r)$ is the Bessel function of first kind, $\alpha_{01}$ is the first zero of $J_0$, $n_{\mathrm{eff}}= 3.2$ is the effective refractive index of the heterostructure; $\Delta n=0.6$ and $\lambda_0$ are the index contrast and central wavelength of the DBRs and $k_m=2\pi n_\mathrm{eff}/\lambda_0$. Figure~\ref{fig:threeExamples} lists $g_{xm}$ for the modes indicated in the density maps. The values of $g_{cm}$ were extracted adapting~\cite{Baker2014,Anguiano2018}, while $g_{cx}$ can be calculated as in \cite{Savona1999,Panzarini1999} (cf. \cite{Note1}). For the near-optimal $\mathcal{I}_g$ values here considered, we notice that the ratio $g_{xm}/\Omega_m\sim 10^{-3}$ is independent of the resonator geometry. Indeed, higher $\Omega_{\mathrm{m}}$ values lead to shorter phonon wavelengths, and larger displacement gradients efficiently activate the deformation potential. As pillars here support the highest phonon frequencies, they present the largest optomechanical coupling ratio $g_{xm}/g_{cm}\sim 10^2$. Related findings for disk and ring resonators are discussed in \cite{Note1}, as a function of the C and M mode indices, showing overall that polaritons experience strongly enhanced optomechanical interactions.
	\begin{figure}[t]
		\centering
		\includegraphics[trim=0cm 0cm 0cm 0cm, width=86mm]{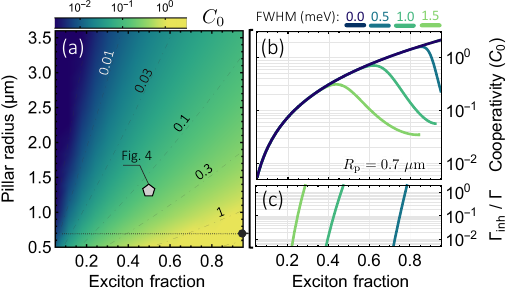}
		\caption{(a) Single-polariton cooperativity versus exciton fraction and pillar radius ($R_p$). (b) Cut through panel (a) for $R_{p}=\SI{0.7}{\um}$ showing the effect of inhomogeneous X broadening (Gaussian linewidth indicated in the inset). (c) Polariton inhomogeneous-broadening $\Gamma_{\mathrm{inh}}$, relative to the mechanical damping $\Gamma$.}
		\label{fig:PillarCoop}
	\end{figure}
	
	As an example, we compute the single polariton cooperativity ($C_0=4 g_{lm}^2/\kappa_l\Gamma$) for the resonator in Fig.~\ref{fig:threeExamples}\,(c) as a function of the pillar radius $R_p$. Deriving the scaling of $x_{\mathrm{ZPF}}$ with $R_p$, and adapting the value of $g_{cm}$ provided in \cite{Anguiano2018}, yields $g_{lm}(R_p)$. The optical decay rate of the heterostructure, including residual absorption at $4\mathrm{K}$~\cite{Sturge1962}, reaches $\kappa_c/2\pi\approx \SI{7.2}{\GHz}$ for 25 DBR pairs (i.e. a quality factor $Q_{c}\sim \num{5e4}$). We also consider a non-radiative exciton decay rate $\kappa_x/2\pi=\SI{4.8}{\GHz}$ \cite{CarlonZambon2020th}, and recall that $\kappa_l=\cos^2\theta_{cx}\kappa_c+\sin^2\theta_{cx}\kappa_x$~\cite{Savona1999}. Due to the co-localization of C and M modes one expects $Q_{m}\approx Q_{c}$; several mechanisms can degrade $Q_{m}$~\cite{Hamoumi2018,Note1}: we take $\Gamma/2\pi=\SI{0.65}{\MHz}$ ($Q_{m}\approx Q_{c}/2$). Given the moderate Q factors at play, we neglect fabrication-induced surface losses. Figure~\ref{fig:PillarCoop}\,(a) shows the cooperativity as a function of the pillar radius and exciton fraction. Interestingly, we observe a region where $C_0\sim 1$. Nevertheless, such a regime is accessible only for large X fractions where detrimental effects related to the matter component become sizable~\cite{Delteil2019,Munoz-Matutano2019}. In particular, the X transition always presents some inhomogeneous broadening. Using the theory developed in \cite{Diniz2011} we show in Fig.~\ref{fig:PillarCoop}\,(b) to which extent this affects $C_0$, while Fig.~\ref{fig:PillarCoop}\,(c) present the polariton inhomogeneous-broadening $\Gamma_{\mathrm{inh}}$ normalized to the mechanical damping $\Gamma$. We have considered a Gaussian broadening with full-width at half-maximum (FWHM) of $0.5$--$\SI{1.5}{\meV}$ and $R_p=\SI{0.7}{\um}$. Coherent control of M requires negligible added phase-noise in the L mode, thus desirably $\Gamma_{\mathrm{inh}}/\Gamma < 1$ \cite{Rabl2009}. Remarkably, Fig.~\ref{fig:PillarCoop}\,(b,c) indicate that state-of-the-art QWs with a broadening below $\SI{0.5}{\meV}$~\cite{Poltavtsev2014}, allow coherent control with $C_0\sim 1$ ($C_q \sim 0.3$ at $\SI{4}{\K}$) for resonators complying with current fabrication technologies. Quantum cavity optomechanics experiments would thus become feasible using few  photons \cite{Galland2014,Fiaschi2021,Fogliano2021} while piezo effects \cite{Fricke1991} may be harnessed to operate such resonators as transducers \cite{Higginbotham2018,Mirhosseini2020,Arnold2020}.

	\textit{Dynamics}~---~Finally, we study how polariton nonlinearities modify dynamical back-action. We consider $g_{cx}/\Omega_m \gg 1$, then $\hat{H}_{lu}$ is off-resonant and the dynamics of the two polariton branches decouples. For a laser detuning $\delta=(\omega-\omega_l)\ll g_{cx}$, we effectively obtain the Hamiltonian of a Kerr resonator coupled to a mechanical mode. In~\cite{Note1}, we derive the quantum Langevin equations (QLEs) ruling the dynamics, calculate the steady-state observables and the regions of dynamical stability in parameter space. Provided single-polariton nonlinearities are weak ($\chi_l/\kappa_l \ll 1$), one can follow the standard linearization approach to study the dynamics of small fluctuations~\cite{Bonifacio1978,Genes2008, Laflamme2011}. Due to the nonlinear term in Eq.~\eqref{eq:EffectiveLPUPHam}, the L fluctuations  $(\delta\hat{\alpha},\delta\hat{\alpha}^{\dag})$ are dynamically coupled. Following~\cite{Asjad2019}, we introduce squeezed displacement operators $\hat{s}=\cosh(r)(\tilde{\alpha}^*\delta\hat{\alpha})+\sinh(r)(\tilde{\alpha}\delta\hat{\alpha}^{\dag})$ where $\tilde{\alpha}=\langle \hat{l}\rangle$, $2r=\mathrm{arctanh}(-\chi\tilde{n}/\tilde{\delta})$, $\tilde{n}=|\tilde{\alpha}|^2$, $\tilde{\delta}=(\omega-\omega_l-2\tilde{\chi}\tilde{n})$ and $\tilde{\chi}=(\chi_l-2g_{lm}^2/\Omega_m)$. The squeezing transformation reduces the QLEs to those of an equivalent harmonic resonator, with a rescaled detuning $\delta_s=\tilde{\delta}\cosh(2r)+\chi_l\tilde{n}\sinh(2r)$, optomechanical coupling $g_s=g_{lm} e^{-r}$ and subject to a squeezed optical bath~\cite{Note1}. Accordingly, the modified mechanical susceptibility adopts the usual expression~\cite{Aspelmeyer2014} upon introducing $(\delta_s,g_s)$, while the displacement spectrum $\overline{S}_{qq}(\omega)$ includes corrections due to correlations in the optical bath~\cite{Note1}.

\begin{figure}[t]
		\centering
		\includegraphics[trim=0cm 0cm 0cm 0cm, width=86mm]{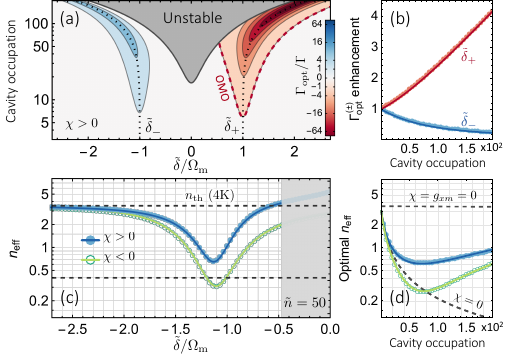}
		\caption{(a) Optomechanical damping rate $\Gamma_{\mathrm{opt}}$ versus detuning ($\tilde{\delta}$), and photon occupation ($\tilde{n}$) for repulsive interactions ($\chi>0$). Dashed black (red) lines trace the sideband detuning $\tilde{\delta}_{\pm}$ (the mechanical-oscillation threshold, OMO); unstable regions are shaded in gray. (b) Enhancement of $\Gamma_{\mathrm{opt}}^{\pm}$ at the sidebands $\tilde{\delta}_{\pm}$ relative to its value for $\chi=0$ as a function of the cavity occupation. (c) Mean phonon occupation $n_{\mathrm{eff}}$ versus detuning for $\tilde{n}=50$. Dashed lines indicate the thermal population and the lowest $n_{\mathrm{eff}}$ for $\chi=0$. (d) Minimal $n_{\mathrm{eff}}$ versus the cavity occupation. In (b,c,d) markers denote numerical solutions while lines show the analytical results~\cite{Note1}.}
		\label{fig:NonlinearSBcool}
	\end{figure}

	As an example, in Fig.~\ref{fig:NonlinearSBcool} we employ this formalism to describe optomechanical amplification and cooling for the pillar indicated by the marker in Fig.~\ref{fig:PillarCoop}(a). According to our previous results, we have $\kappa_l/2\pi=\SI{6.5}{\GHz}$, $\Omega_m/\kappa_l=3$, $\Gamma/\kappa_l=10^{-4}$, $g_{lm}/\kappa_l=0.002$, $\chi_l/\kappa_l=0.03$ (cf.~\cite{Note1}), yielding a cooperativity $C_0=0.15$. We consider the system to be pre-cooled to $4~\mathrm{K}$. Figure~\ref{fig:NonlinearSBcool}\,(a) presents the optomechanical damping rate $\Gamma_{\mathrm{opt}}$ as a function of the effective laser detuning from the L resonance ($\tilde{\delta}$), and of the polariton occupation; regions of single-mode instability are shaded in gray. The optomechanical self-oscillation threshold (OMO) is indicated with a dashed red line. We can observe two main differences with respect to the harmonic resonator case ($\chi=0$). First, the sidebands positions $\tilde{\delta}_{\pm}$ depend on the polariton density as $\tilde{\delta}_{\pm}\approx\pm\sqrt{\Omega_m^2+\tilde{\chi}^2\tilde{n}^2}$ (assuming $\Omega_m/\kappa_l\gg1$). Second, the extremal values of the optomechanical damping, denoted $\Gamma_{\mathrm{opt}}^{\pm}$ are asymmetric: for repulsive $\chi>0$ (attractive $\chi<0$) nonlinearities the Stokes sideband is suppressed (enhanced) with respect to its value in absence of nonlinearities; the opposite holds for the anti-Stokes sideband. Repulsive interactions boost the optomechanical gain, resulting in efficient ultra-low threshold OMOs (here as low as $\tilde{n}\approx7$). Quantitatively, assuming $\Gamma_{\mathrm{opt}}(\omega)\approx\Gamma_{\mathrm{opt}}(\Omega_m)$, yields the sideband enhancement factor $\eta_{\pm}=(1+\chi\tilde{n}/\tilde{\delta}_{\pm})^{1/2}/(1+\chi\tilde{n}/\tilde{\delta}_{\mp})^{1/2}$, traced with solid lines in Fig.~\ref{fig:NonlinearSBcool}\,(b) versus cavity occupation for $\chi_l>0$; markers indicate the results obtained by exact diagonalization of the QLEs~\cite{Note1}.
	
	Concerning the cooling performance, we start by noticing that the mean effective phonon occupation in the system ($n_{\mathrm{eff}}$) is related to the internal energy of the oscillator $\Omega_m(n_{\mathrm{eff}}+1/2)=\int \frac{\mathrm{d}\omega}{4\pi} \tilde{m}(\Omega^2 +\omega^2)\overline{S}_{qq}(\omega)$~\cite{Genes2008}. In Fig.~\ref{fig:NonlinearSBcool}\,(c) we trace $n_{\mathrm{eff}}$ as a function of laser detuning for $\tilde{n}=50$, both for the case of repulsive (solid markers) and equal but attractive (hollow markers) nonlinearities. In both cases occupations below unity can be achieved. As the Stokes sideband is reduced (enhanced) when $\chi>0$ ($\chi<0$) with respect to the linear case ($\chi=0$), one would expect cooling protocols to be more efficient for $\chi<0$. In Fig.~\ref{fig:NonlinearSBcool}\,(e) we trace the minimum achievable $n_{\mathrm{eff}}$ versus cavity occupation for the three cases, showing this is not generally true. For large interaction energies $\tilde{\chi} \tilde{n} \sim \kappa$, even if the sideband enhancement factor $\eta_{-}$ becomes large, the sideband peak detunes from $\Omega_m$ (unaffected by optical nonlinearities) thus suppressing the scattering rate by $\sim\kappa_l^2[4(\Omega_m-\tilde{\delta}_{-})^2+\kappa_l^2]^{-1}$. For the specific resonator  considered in Fig.~\ref{fig:NonlinearSBcool}, these two opposing effects result in a finite yet modest improvement of the cooling performance. Nevertheless, our analytic results indicate that the cooling enhancement becomes large in the bad cavity limit, see \cite{Note1,Laflamme2011,Zoepfl2022}. Furthermore, to ease the comparison with the linear case, we kept $g_{xm}$ constant. In practice, without excitons, $\chi_l=g_{xm}=0$ and a $(g_{lm}/g_{cm})^2\sim 10^3$ higher cavity occupation is required to reach $n_{\mathrm{eff}}\sim 1$, see Fig.~\ref{fig:NonlinearSBcool}\,(d). Finally, we notice that for large cavity occupations ($\tilde{n}\sim 10^3$), one enters the strong optomechanical coupling regime~\cite{Note1,Yeo2014,Montinaro2014}, characterized by hybrid M-C-X quasiparticles, or phonoritons, akin to those predicted for cavities embedding hBN flakes~\cite{Latini2021}, or two-level atoms~\cite{Restrepo2017}.
	
	\textit{Outlook}~---~Our results demonstrate the potential of harnessing QW exciton polaritons to enhance optomechanical interactions and indicate that a near-unity single-polariton cooperativity can be achieved in state-of-the-art resonators. Contextually, we adapted the theory of dynamical back-action to include polariton interactions and showed that sideband cooling at $4\mathrm{K}$ is sufficient for ground-state preparation. We foresee that stronger nonlinearities~\cite{Delteil2019,Munoz-Matutano2019} could be exploited to stabilize non-classical mechanical states~\cite{Ma2021}. Our analysis can be readily extended to multi-mode scenarios, naturally emerging in coupled microresonator arrays, where a simultaneous engineering of the polariton and phonon dispersion would disclose a variety of applications~\cite{Peano2015,Nielsen2017,Ruesink2018}.
	
	\begin{acknowledgments}
		This work was supported by the MaCaCQu Flagship project of the Paris Saclay Labex (ANR-10-LABX-0035), by ANR via the project UNIQ, by the H2020-FETFLAG project PhoQus (820392), by the QUANTERA project Interpol (ANRQUAN-0003-05), and by the European Research Council via the project ARQADIA (949730), and the Consolidator grant NOMLI (770933). We thank Daniel Lanzillotti-Kimura and Philippe St-Jean for valuable discussions as well as Jérémy Bon for numerical assistance.
	\end{acknowledgments}

\bibliographystyle{apsrev4-1}
\bibliography{biblioLMS.bib}

\clearpage

\vspace{1cm}
\renewcommand{\thefigure}{S\arabic{figure}}
\renewcommand{\theequation}{S\arabic{equation}}
\setcounter{equation}{0}
\setcounter{figure}{0}

\begin{center}
	\large{\textsc{Supplementary materials}}
\end{center}

\appendix

\subsection{A - Derivation of the electromechanical coupling\label{app:A}}

We provide details about the derivation of the electromechanical coupling in hybrid resonators comprising a quantum well, Eq.~\eqref{eq:approxGxm} in the main text.

The coupling between excitons and phonons originates from modifications of the semiconductor band structure induced by strain in the material~\cite{Bardeen1950,Anselm1955,Herring1956,Bir-Pikus1974}. The resulting energy shift for inter-band transitions is captured by the deformation potential~\cite{Piermarocchi1996}
\begin{equation}\label{eq:A.1}
	U(\vb*{r}_e,\vb{r}_h) = a_e \Sigma_n(\vb*{r}_e) - a_h \Sigma_n(\vb*{r}_h),
\end{equation}
where $a_e$ and $a_h$ are the electron and hole deformation potentials, and $\Sigma_n(\vb*{r}) = \nabla \cdot \vb*{u}_n(\vb*{r})$ denotes the mechanical strain at $\vb*{r}$ imputable to the presence of a phonon in the mechanical mode $n$ under consideration. Here, $\vb*{u}_n(\vb*{r}) = x_{n}^{\mathrm{ZPF}}\phi_n^\pidx{m}(\vb*{r})$ denotes the corresponding displacement, where $x_{n}^{\mathrm{ZPF}} = \sqrt{\hbar/2M_n\Omega_n}$ is the magnitude of the zero-point fluctuations of the mechanical degree of freedom, of mass $M_m$ and angular frequency $\Omega_n$, and $\phi_n^\pidx{m}$ its associated wavefunction, normalized as $\int_V\mathrm{d}\vb*{r}\lvert\phi_n^\pidx{m}(\vb*{r})\rvert^2 = V$. The translational invariance of the system allows separating the QW exciton center-of-mass and relative degrees of freedom. One can then neglect the exciton dispersion---the exciton effective mass being roughly four orders of magnitude larger than the one of cavity photons for GaAs quantum wells---and expand the exciton center-of-mass wavefunction in the same in-plane basis as that of the optical modes of the cavity. Focusing on some specific optical mode $\lambda$, as described by a wavefunction of the form $\phi_{\lambda}^{\pidx{c}}(\vb*{r}) = F_\lambda(\vb*{r}_\parallel)f_\lambda(z)$, with $\int_S\mathrm{d}\vb*{r}_\parallel \lvert F_\lambda(\vb*{r}_\parallel)\rvert^2 = 1$, the relevant exciton wavefunction is
% (\vb*{R}_\parallel,\vb*{\rho}_\parallel,z_h,z_e)
\begin{equation}
	\phi_{\lambda}^{\pidx{x}}(\vb*{r}_e,\vb*{r}_h) = F_\lambda(\vb*{R}_\parallel)\phi(\rho_\parallel)\chi_h(z_h)\chi_e(z_e),
\end{equation}
where $\vb*{R}_\parallel = \frac{m_e\vb*{r}_{e,\parallel} + m_h\vb*{r}_{h,\parallel}}{m_e + m_h}$ and $\vb*{\rho}_\parallel = \vb*{r}_{e,\parallel} - \vb*{r}_{h,\parallel}$ denote the center-of-mass and relative coordinates of the carriers.

In the low density regime, one may assume a bosonic statistics for the excitons~\cite{Carusotto2013} and  express the deformation potential introduced in Eq.~\eqref{eq:A.1} in second quantization by means of the usual prescription~\cite{Altland2010}
\begin{equation}
	\hat{H}_{xm} = -\hbar g_{xm}^{\lambda,n}\hat{d}_\lambda^\dagger\hat{d}_\lambda\hat{b}_n + \mathrm{H.c.},
\end{equation}
with the exciton-phonon coupling factor:
\begin{widetext}
	\begin{align}\label{eq:exactGxm}
		-\hbar g_{xm}^{\lambda,n} &:= \bra{0,\phi_{\lambda}^{\pidx{x}}}\hat{U}\ket{\phi_{n}^{\pidx{m}},\phi_{\lambda}^{\pidx{x}}} = \int_V\mathrm{d}\vb*{r}_e\mathrm{d}\vb*{r}_h\phi_\lambda^{\pidx{x}\star}(\vb*{r}_e,\vb*{r}_h)U(\vb*{r}_e,\vb{r}_h)\phi_\lambda^\pidx{x}(\vb*{r}_e,\vb*{r}_h)\nonumber\\
		=&\int\mathrm{d}\vb*{R}_\parallel\mathrm{d}\vb*{\rho}_\parallel\mathrm{d}z \lvert F_\lambda(\vb*{R}_\parallel)\rvert^2\lvert \phi(\vb*{\rho}_\parallel)\rvert^2\Bigl(a_e\Sigma_n(\vb*{R}_\parallel+\tfrac{m_h}{M}\vb*{\rho}_\parallel+z\vb*{e}_z)\lvert\chi_e(z)\rvert^2 - a_h\Sigma_n(\vb*{R}_\parallel+\tfrac{m_e}{M}\vb*{\rho}_\parallel+z\vb*{e}_z)\lvert\chi_h(z)\rvert^2\Bigr).
	\end{align}
\end{widetext}

This expression can be greatly simplified by neglecting the exciton Bohr radius over the mechanical in-plane wavelength and the thickness of the QW over the typical out-of-plane variations of the strain. The exciton-phonon coupling becomes then solely associated the overlap between the exciton envelope and the single-phonon strain at the location of the well:
\begin{equation}\label{eq:SMapproxGxm}
	\hbar g_{xm}^{\lambda,n} \approx -(a_e - a_h)\int_S\mathrm{d}\vb*{R}_\parallel\lvert F_\lambda(\vb*{R}_\parallel)\rvert^2\Sigma_n(\vb*{R}_\parallel, z_\mathrm{QW}).
\end{equation}

In the following, we derive explicit expression for the exciton wavefunction in the case of shallow QWs, and for the optical and strain field envelopes for three relevant resonator geometries: microdisk, microring and micropillar. This will allow us to benchmark the validity of Eq.~\eqref{eq:SMapproxGxm} against the exact numerical results descending from Eq.~\eqref{eq:exactGxm}. 

\subsection{B - Shallow QW excitons\label{app:B}}	

\begin{figure}[tb]
	\centering
	\includegraphics[trim=0cm 0cm 0cm 0cm, width=86mm]{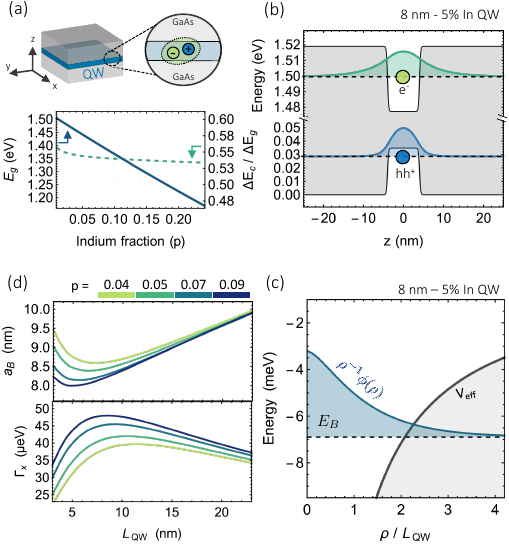}
	\caption{Excitons in shallow QWs: (a) Graphic representation of a QW for electrons and holes formed by a few-nanometer thick layer of $\mathrm{In_{p}Ga_{1-p}As}$ embedded in a host GaAs matrix. Below we report the dependence on the molar Indium fraction (p) of the bandgap ($E_g$, solid line) and of the conduction band mismatch ($\Delta E_c$, dashed line) at the heterointerface. (b) Calculated electron and heavy-hole envelopes along the QW growth axis ($z$) for a $8~\mathrm{nm}$ thick $\mathrm{In_{0.05}Ga_{0.95}As}$ QW. (c) Corresponding profiles of the pseudo-potential $V_{\mathrm{eff}}$ and of the first eigenmode $\phi(\rho)$ of the radial exciton problem, yielding the binding energy $E_B\approx -6.8~\mathrm{meV}$. (d) Exciton Bohr radius ($a_B$) and radiative   half-linewidth ($\Gamma_x$) as a function of the QW thickness ($L_\mathrm{QW}$) and Indium concentration obtained via the pseudo-potential method.}
	\label{fig:ShallowQWs_Excitons}
\end{figure}

Shallow $\mathrm{In_{p}Ga_{1-p}As}$ quantum-wells embedded in a GaAs matrix are particularly relevant to our study as they are both compatible with the epitaxial growth of $\mathrm{AlGaAs}$-based heterostructures, and present nearly vanishing inhomogeneous broadening~\cite{Poltavtsev2014}. As mentioned in the main text, the latter is a key figure of merit in order to access the large single phonon-polariton cooperativity limit while keeping a coherent control over the system. In the following we summarize the method we used to determine the exciton wavefunction and other key parameters, as the Bohr radius $a_B$ and radiative exciton linewidth, ultimately determining the light-matter coupling $g_{cx}$. Hereafter we consider a single QW lying parallel to the $xy$ plane, and characterized by a thickness $L_\mathrm{QW}$, see the sketch in Fig.~\ref{fig:ShallowQWs_Excitons}\,(a). In the QW layer, the bandgap energy $E_g$ is locally lowered with respect to GaAs, because of the presence of InAs in the alloy. The scaling of $E_g$ depends on the relative In molar fraction $p$ and is traced as a solid line in Fig.~\ref{fig:ShallowQWs_Excitons}\,(a) (adapted from \cite{Paul1991}). If we denote $\Delta E_g$ the bandgap offset along the growth axis $z$, the offset in the valence and conduction bands at the $\Gamma$ point satisfies $\Delta E_g = \Delta E_v + \Delta E_c$, with relative weights determined by the matching of the Fermi energies at the heterointerface. In Fig.~\ref{fig:ShallowQWs_Excitons}\,(a) we report $\Delta E_c/\Delta E_g$ as a function of the indium content (dashed line, adapted from \cite{Zubkov2004}). 

The standard approach to the electron-hole in a QW problem, that is the separability of the Hamiltonian with respect to the radial and transverse coordinates, results inaccurate to describe shallow QW excitons. Indeed, in order to neglect the dependence on the relative  electron-hole distance along the transverse coordinate of the Coulomb potential, two conditions need to be satisfied: the envelopes $\chi_{e,h}(z_{e,h})$ of the electron and hole along $z$ must be nearly identical, and the quantum well width ($L_\mathrm{QW}$) must be much smaller than the exciton Bohr radius ($a_B$)~\cite{Bastard1992}. However, in shallow QWs, the confinement energy for carriers becomes comparable with the band offsets: the electron and hole wavefunction spread in the GaAs matrix with significantly different penetration depths, being the electron effective mass typically one an order of magnitude smaller than the one of the heavy-hole~\cite{Levinshtein1996}. This is quite evident in Fig.~\ref{fig:ShallowQWs_Excitons}\,(b), showing the solutions of the particle in a box problem for the electrons and heavy-holes in $\SI{8}{\nm}$ wide $\mathrm{In_{0.05}Ga_{0.95}As}$ QW; the effective masses of the carriers were taken from \cite{Adachi1982}.

An alternative approach that effectively restores the separability of the problem relies on the definition of a pseudo-potential for the relative in plane motion of electron and holes~\cite{Moore1990}. The first step consists in determining the electron-hole envelopes $\chi_{e,h}(z_{e,h})$ in absence of the Coulomb interaction, thus defining the pseudo-potential
\begin{equation}
	V_{\mathrm{eff}}(\rho)=-\frac{e^2}{4\pi \epsilon_0 \epsilon}\int \mathrm{d}z_e \mathrm{d}z_h \frac{|\chi_{e}(z_{e})|^2|\chi_{h}(z_{h})|^2}{\sqrt{\rho^2+(z_e-z_h)^2}},
\end{equation}
where $\rho$ is the in-plane distance between the electron and the hole, $e$ is the electron charge and $\epsilon\epsilon_0$ is the dielectric permittivity of GaAs. As $z_{e,h}$ are  marginalized by the integral, one then needs to solve only the radial problem associated to $V_{\mathrm{eff}}$ in order to determine the exciton radial envelope $\phi(\rho)$ and binding energy $E_b$. Once $E_b$ is determined, one can effectively include the Coulomb interaction when determining the transverse envelopes $\chi_{e,h}$ and iterate self-consistently the procedure until convergence. As an example, in Fig.~\ref{fig:ShallowQWs_Excitons}\,(c) we trace the pseudo potential and the radial profile of $\phi(\rho)$ calculated for a $\SI{8}{\nm}$ wide $\mathrm{In_{0.05}Ga_{0.95}As}$ QW; the dashed black line indicates the exciton binding energy. We repeated the calculation for different Indium contents and as function of the QW width. Knowing the electron-hole and exciton envelopes and we could then deduce the exciton Bohr radius $a_B=\int \mathrm{d}\rho \phi(\rho) / \int \mathrm{d}\rho \phi(\rho)/\rho$ and the radiative exciton lifetime $\tau_x=\hbar/\Gamma_x$, cf.~\cite{Andreani1995}. We summarize these results in Fig.~\ref{fig:ShallowQWs_Excitons}\,(d); each solid curve corresponds to a different Indium content, as indicated by the legend.

\subsection{C - Optical and vibrational modes: Microdisk\label{app:C}}

Concave resonators exhibit normal modes travelling at their inner periphery, bearing the name of whispering gallery modes (WGMs) and first described by Lord Rayleigh~\cite{Strutt1878,Rayleigh1910,Rayleigh1914}. Here, we describe the (in-plane) TE-polarized WGMs of a semiconducting disk of radius $R_d$ and thickness $L_z$.

The electromagnetic energy density within the disk reads
\begin{equation}
	\mathcal{U}_c = \frac{1}{2}\bigl(\epsilon\lVert\vb*{E}\rVert^2 + \frac{1}{\mu}\lVert\vb*{B}\rVert^2\bigr),
\end{equation}
where $\epsilon$ and $\mu$ denote the permittivity and the permeability of the material. For TE modes, one has $A_z = 0$ and thus $\vb*{A} = \sum_{\sigma=\pm} A_\sigma \vb*{e}_\sigma$, in terms of circularly polarized components $A_\pm$ and their associated rotating unit vectors $\vb*{e}_\pm = e^{\pm i\theta}(\vb*{e}_r \pm i\vb*{e}_\theta)/\sqrt{2}$. In the following we adopt the effective-refractive-index approach, dimensionally reducing the problem to the field's bare in-plane dependences, accounting for its vertical confinement by means of an effective refractive index $n_\mathrm{eff}$ depending on the thickness $L_z$ of the disk. The classical energy stored in the cavity is hence, for the component under consideration, given by
\begin{equation}\label{eq:23}
	H_c = \int_V\mathrm{d}\vb*{r}\Bigl\lbrace\frac{1}{2}\epsilon \dot{A}_\sigma^2 + \frac{1}{2}\epsilon c^2 A_\sigma \Bigl(\frac{1}{r^2}\partial_\theta^2 + \frac{1}{r}\partial_r[r]\partial_r\Bigl)A_\sigma\Bigr\rbrace,
\end{equation}
where $c = c_0/n_{\mathrm{eff}}$ denotes the speed of light in the medium. This Hamiltonian may be diagonalized in the eigenbasis of the solutions of the following Helmholtz equation:
\begin{equation}
	\ddot{A}_\sigma = -c^2\Bigl(\frac{1}{r^2}\partial_\theta^2 + \frac{1}{r}\partial_r[r]\partial_r\Bigl)A_\sigma.
\end{equation}
By separation of variables, this can be split into a harmonic equation for the azimuthal dependence and a Bessel equation for the radial one, yielding the following set of base wavefunctions
\begin{equation}\label{eq:1.6}
	\phi_{p,\ell}^\pidx{c}(r,\theta,z) = N_{p,\ell} f(k_z z)J_\ell(k_{p,\ell}r)\cos(\ell\theta),
\end{equation}
where complete reflection of the electromagnetic field at the radial boundaries of the material was assumed for simplicity. These satisfy $\ddot{\phi}_{p,\ell}^\pidx{c} = -c^2k_{p,\ell}^2 \phi_{p,\ell}^\pidx{c}$, where the wavefunction $k_{p,\ell}$ can be expressed as a function of the $p$th root of the $\ell$th Bessel function of the first kind $\alpha_{\ell,p}$ simply as $k_{p,\ell} = \alpha_{\ell,p}/R_d$. Rather unsurprisingly, we recover exactly Rayleigh's solution~\cite{Rayleigh1910}. The normalization is given by
\begin{equation}
	N_{p,\ell} = \frac{\sqrt{2/V}}{\lvert J_{\ell+1}(k_{p,\ell}R_d)\rvert}\Bigl(\int\frac{\mathrm{d}z}{L_z}\lvert f(k_z z)\rvert^2\Bigr)^{-1/2},
\end{equation}
such that wavefunctions satisfy the orthonormalization condition $\int_V\mathrm{d}\vb*{r}\phi_{p\ell}^{\pidx{c}\star}(\vb*{r}) \phi_{p'\ell'}^\pidx{c}(\vb*{r}) = \delta_{p,p'}\delta_{\ell,\ell'}$.

The vertical dependence $f$ is a solution of the following wave equation:
\begin{equation}\label{eq:WaveEqZ}
	\partial_z^2 f(k_z z) + k_0^2\bigl(n(z) - n_{\mathrm{eff}}^2\bigr)f(k_z z) = 0,
\end{equation}
where $k_0 = 2\pi/\lambda_0$, $n(z) = n$ within the material and $n(z) = 0$ outside of it. We shall only consider the first even solution to Eq.~\eqref{eq:WaveEqZ}, as given by:
\begin{equation}
	f(k_z z) = A \begin{cases}
		\cos(k_z z), & \lvert z\rvert \leq L_z/2;\\
		\cos(\frac{k_z L_z}{2}) e^{-\alpha k_z(\lvert z\rvert - L_z/2)}, & \text{else};
	\end{cases}
\end{equation}
with
\begin{equation}
	k_z = k_0\sqrt{n^2 - n_\mathrm{eff}}, \quad \alpha = \sqrt{\frac{n_\mathrm{eff} - 1}{n^2 - n_\mathrm{eff}}},
\end{equation}
and $n_\mathrm{eff}$ the largest root of
\begin{equation}
	\tan(\frac{k_z L_z}{2}) - \alpha = 0.
\end{equation}
This is shown in Fig.~\ref{fig:neffLeffRabi}(a) as a function of the vertical confinement.

Now, by expanding the vector potential into the WGM basis, $A_\sigma(r,\theta,z) = \sum_{p = 1,\ell = 0}^{+\infty} A_{p,\ell}\phi_{p,\ell}^\pidx{c}(r,\theta,z)$, and defining the conjugated generalized coordinates $u_{p,\ell} = \sqrt{\epsilon}A_{p,\ell}$ and $\pi_{p,\ell} = \dot{u}_{p,\ell}$, the Hamiltonian reduces to that of a set of independent harmonic oscillators
\begin{equation}
	H_c = \sum_{p,\ell}\Bigl(\frac{1}{2}\pi_{p,\ell}^2 + \frac{1}{2}\omega_{p,\ell}u_{p,\ell}^2\Bigr),
\end{equation}
with $\omega_{p,\ell} = c k_{p,\ell}$. Adopting the canonical quantization prescriptions $(u_{p,\ell}, \pi_{p,\ell}) \mapsto (\hat{u}_{p,\ell}, \hat{\pi}_{p,\ell})$, with operators $\hat{u}_{p,\ell}$ and $\hat{\pi}_{p,\ell}$ satisfying the commutation relation $[\hat{u}_{p,\ell}, \hat{\pi}_{p',\ell'}] = i\hbar\delta_{p,p'}\delta_{\ell,\ell'}$, and introducing the usual photon annihilation operators,
\begin{equation}
	\hat{a}_{p,\ell} = \sqrt{\tfrac{\omega_{p,\ell}}{2\hbar}}\hat{u}_{p,\ell} + i \sqrt{\tfrac{1}{2\hbar\omega_{p,\ell}}}\hat{\pi}_{p,\ell},
\end{equation}
one obtains the final Hamiltonian of the WGMs' photons:
\begin{equation}
	\hat{H}_c = \sum_{p,\ell}\hbar\omega_{p,\ell}(\hat{a}_{p,\ell}^\dagger\hat{a}_{p,\ell}^{\mathstrut} + 1/2).
\end{equation}
This, altogether with the direct-space representation of the vector potential,
\begin{equation}\label{eq:30}
	\vb*{\hat{A}}(\vb*{r}) = \sum_{p,\ell}\phi_{p,\ell}^\pidx{c}(\vb*{r})\sqrt{\frac{\hbar}{2\epsilon \omega_{p,\ell}}}(\hat{a}_{p,\ell} + \hat{a}_{p,\ell}^\dagger)\vb*{e}_\sigma,
\end{equation}
provides a complete quantum description of the disk's WGMs.

\begin{figure}[tb!]
	\centering
	\includegraphics[width=.9\linewidth]{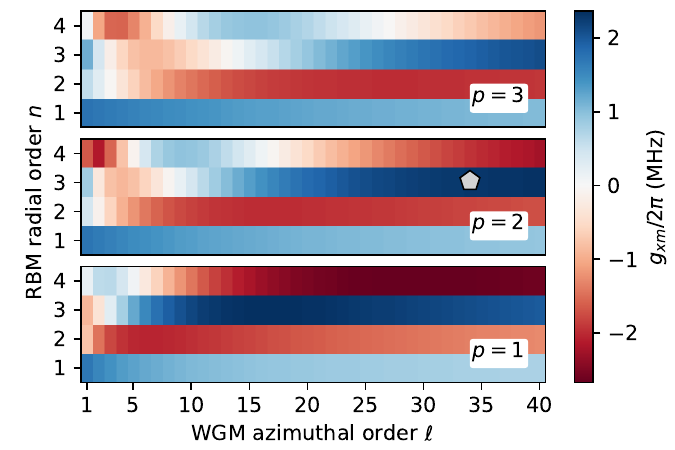}
	\caption{Electromechanical coupling for a disk of radius $R_d = \SI{2}{\um}$ and thickness $L_z = \SI{200}{\nm}$ bearing a mechanical radial breathing mode of order $n$ and a whispering gallery mode characterized by a principal and azimuthal orders $(p,\ell)$. The value $g_{xm}/2\pi = \SI{2.29}{\MHz}$ considered in Fig.~\ref{fig:threeExamples} of the main text is highlighted with a marker.}
	\label{fig:gxmDisk}
\end{figure}

Let us now consider the mechanical degrees of freedom of the semiconductor disk. We will assume the disk material to be homogeneous and isotropic, as characterized by the two Lamé parameters $(\lambda,\mu)$ and its volumetric mass $\rho$.
Any perturbation of its radial or vertical profiles results in internal stresses that counteract the deformation and tend to restore equilibrium. Inside a continuous elastic material, these conservative forces follow from the Hooke's potential energy density~\cite[Chapter~2]{Landau2008}:
\begin{equation}\label{eq:C.26}
	\mathcal{U}_m = \frac{1}{2}\sigma^{ik}\varepsilon_{ik} = \frac{1}{2}\lambda \varepsilon^{ii}\varepsilon_{ii} + \mu \varepsilon^{ik}\varepsilon_{ik},
\end{equation}
where $\sigma_{ik}$ and $\varepsilon_{ik}$ denote the stress and strain tensors, respectively, and where summation over repeated indices is implicitly assumed. We will consider cylindrical coordinates $(r, \theta, z)$. Any in-plane stress within the disk results in a sizable strain $\varepsilon_{zz}$ along the out-of-plane direction because of the finite Poisson ratio \smash{$\nu = \frac{\lambda}{2(\lambda + \mu)}$} of the material. This difficulty can be circumvented by adopting the so-called plane-stress condition for the disk, routinely used when addressing thin plates. Under this assumption, the state of the disk is such that $\sigma_{zz} = 0$, that is, the vertical strain adapts to the presence of internal in-plane stresses everywhere within the disk's volume. Imposing this constraint on Eq.~\eqref{eq:C.26}, one is left with an identical expression for a reduced 2D problem, upon introducing modified first Lamé parameter:
\begin{equation}
	\mathcal{U}_m = \frac{1}{2}\tilde{\lambda} \varepsilon^{ii}\varepsilon_{ii} + \mu \varepsilon^{ik}\varepsilon_{ik}\;\text{(2D)},\quad
	\tilde{\lambda}=\frac{2\mu}{\lambda+2\mu}\lambda.
\end{equation}
Accordingly,
\begin{equation}
	\mathcal{U}_r = \frac{\tilde{\lambda} + 2\mu}{2}\bigl((\partial_r u_r)^2 + (u_r/r)^2\bigr).
\end{equation}
By further introducing the density of kinetic energy $\mathcal{T}_r = \frac{1}{2}\rho\dot{u}_r^2$ associated to this degree of freedom, the RBM motion is described by the following classical Lagrangian
\begin{equation}
	L_r = \int_V\mathrm{d}\vb*{r}\Bigl\lbrace\frac{1}{2}\rho\dot{u}_r^2 - \frac{1}{2}\rho c_s^2 u_r \Bigl(\frac{1}{r}\partial_r[r]\partial_r - \frac{1}{r^2}\Bigr)u_r\Bigr\rbrace,
\end{equation}
where
\begin{equation}
	c_s = \sqrt{\frac{E}{\rho(1-\nu^2)}}
\end{equation}
is the propagation speed of longitudinal acoustic waves in the material, expressed in terms of Young's modulus $E$ and Poisson's ratio $\nu$. The Lagrangian becomes diagonal in the basis of first order Bessel functions of the first kind
\begin{gather}
	\phi_n^\pidx{m} = \mathcal{N}_n J_1(K_n r),\\
	\mathcal{N}_n^{-1} = \sqrt{J_1(K_n R_d)^2 - J_{0}(K_n R_d) J_{2}(K_n R_d)},\nonumber
\end{gather}
where the normalization was chosen such that $\int_V\mathrm{d}\vb*{r} \phi_{n}^\pidx{m}(r)\phi_{m}^\pidx{m}(r) = V \delta_{n,m}$. The eigenbasis of the Lagrangian depends upon the choice of boundary conditions. Here, for the modes to correspond to stationary vibrational states of the resonating material, the mechanical wave vectors $K_n$ must be such that the in-plane radial stress induced by the deformation of the material, as given by $\sigma_{rr} = \tilde{\lambda}(\varepsilon_{rr} + \varepsilon_{\theta\theta}) + 2\mu\varepsilon_{rr} = (\tilde{\lambda} + 2\mu)\partial_r u_r + \tilde{\lambda}u_r/R$, vanish at the boundaries of the disk. The $n$th mechanical wave vector is thus given by the $n$th finite root of
\begin{equation}
	K_n R_d J_0(K_n R_d) - (1-\nu)J_1(K_n R_d) = 0,
\end{equation}
solely depending on the Poisson ratio of the disk's material. We here give the wave vectors of the first three RBMs for GaAs ($\nu = 0.31$):

\begin{table}[h!]
	%\caption{\label{tab:table2}%
	%    A table that fits into a single column of a...
	%}
	\begin{tabular}{@{\qquad} c @{\qquad} c @{\qquad} c @{\qquad}}
		\toprule
		$K_1 R_d$ & $K_2 R_d$ & $K_3 R_d$\\
		\colrule
		2.055 & 5.391 & 8.573\\
		\botrule
	\end{tabular}
\end{table}

By Fourier-Bessel expanding the radial displacement, $u_r(r) = \sum_{n} u_n \phi_n^\pidx{m}(r)$, introducing the conjugated generalized moment $\pi_n = \partial L_r/\partial \dot{u}_n = M_m \dot{u}_n$, with $M_m = \rho V$, and Legendre-transforming the diagonalized Lagrangian, the Hamiltonian reduces to that of a set of independent harmonic oscillators:
\begin{equation}
	H_m = \sum_n\Bigl\lbrace\frac{1}{2M_m}\pi_n^2 + \frac{1}{2}M_m\Omega_n^2 u_n^2\Bigr\rbrace,
\end{equation}
with the mechanical angular frequency $\Omega_n = c_s K_n$.
Applying the canonical quantization prescription $(u_{n}, \pi_{n}) \mapsto (\hat{u}_{n}, \hat{\pi}_{n})$, with operators $\hat{u}_{n}$ and $\hat{\pi}_{n}$ satisfying the canonical commutation relation $[\hat{u}_{n}, \hat{\pi}_{n'}] = i\hbar\delta_{n,n'}$, and introducing the usual phonon annihilation operators,
\begin{equation}
	\hat{b}_n = \sqrt{\frac{M_m\Omega_n}{\hbar}}\hat{u}_n + i\sqrt{\frac{\Omega_n}{\hbar M_m}}\hat{\pi}_n,
\end{equation}
one obtains the final Hamiltonian of the RBMs' phonons:
\begin{equation}
	\hat{H}_m = \sum_{n}\hbar\Omega_n(\hat{b}_n^\dagger\hat{b}_n^{\mathstrut} + 1/2).
\end{equation}
This, altogether with the direct-space representation of the mechanical displacement,
\begin{equation}
	\hat{\vb*{u}}(r) = \sum_{n}\phi_n^\pidx{m}(r)x_n^\mathrm{ZPF}(\hat{b}_{n} + \hat{b}_n^\dagger)\vb*{e}_r,
\end{equation}
furnishes a complete quantum description of the disk RBMs' motion. Here, $x_{n}^{\mathrm{ZPF}} = \sqrt{\hbar/2M_m\Omega_n}$ is the length-scale of zero-point fluctuations, as given by $x_{n}^{\mathrm{ZPF}} = \sqrt{\expval{\hat{u}_n^2}{0}}$.

These expressions were used to evaluate the electromechanical coupling strength of the disk in Fig.~\ref{fig:threeExamples} of the main text. In Fig.~\ref{fig:gxmDisk}, we show the value of this parameter and its dependence on the considered WGM indices for a disk of radius $R_d = \SI{2}{\um}$ and thickness $L_z = \SI{200}{\nm}$.

In the above description, the optical and mechanical modes were considered as completely isolated from the outside world. In practice, however, dissipative processes impact their quality factors.

Optical losses in disk resonators are well understood~\cite{Parrain2014}. % The relevant ref is here 10.1364/OE.23.019656
These are of various origins. First, bending losses due to incomplete internal reflection of the electromagnetic field under strong spatial confinement may degrade the quality factor. While this effect strongly depends on the geometry of the resonator and the considered optical mode, for the considered thickness $L_z = \SI{200}{\nm}$, this contribution was found of subleading order in disks of radius larger than $R_d = \SI{1.5}{\um}$ for modes at $\lambda = \SI{1550}{\nm}$, and should be all the more so at the shorter wavelengths considered herein. Light scattering due to irregularities at the lateral boundaries of the disks incurs in additional intrinsic losses, bounding the optical quality factors by roughly a million. Finally, residual linear absorption, due to the presence of states within the gap stemming from the surface of the material, accounts for most of the optical linewidth. This reduces the typical intrinsic optical quality factor to $Q_c \gtrsim 10^5$, although this may be greatly mitigated by surface passivation, yielding quality factors above a million.% (10.1364/OPTICA.4.000218)

On the mechanical side, radial breathing modes dissipate energy into the substrate via the disk pedestal. These clamping losses may be analytically assessed within the effective two-dimensional approach assumed above~\cite{Hao2007} and mitigated by reducing the pedestal section, at the cost of making the nanofabrication more involved. Resonating disks also dissipate energy into the surrounding fluid through fluidic damping processes, although these become negligible when working in high vacuum. %(Theory and experiments in 10.1038/nnano.2015.160)
Finally, intrinsic properties of the material further impact the mechanical linewidth. These, thoroughly studied in Ref.~\cite{Hamoumi2018}, include visco-elastic and thermo-elastic effects, as well as the presence of microscopic defects in the material, acting as relaxing two-level systems that couple to the acoustic phonons of the radial breathing modes of interest. The former two vanish at temperatures below $T=\SI{25}{\K}$ and are dominated by the latter at all temperatures. This sets a size-independent lower bound on the attainable mechanical linewidth around $\Gamma_m/2\pi \sim \SI{15}{\kHz}$. The freezing of the two-level systems, though, predicted at temperatures lower than $T=\SI{10}{\milli\K}$~\cite{Hamoumi2018}, should in principle allow to reach quality factors in excess of $Q_m = 10^6$ at $T=\SI{100}{\milli\K}$ in this platform.

\subsection{D - Optical and vibrational modes: Microring\label{app:D}}

We shall now consider an annular disk of inner and outer radii $R_i$ and $R_d$ and thickness $L_z$. The optical Hamiltonian of the ring has the same form as the one of the disk [Eq.~\eqref{eq:23}], but distinct boundary conditions and  associated photonic wave functions: 
\begin{align}
	\phi_{p,\ell}^\pidx{c}(r,\theta,z) &= N_{p,\ell} f_q(k_q z)R_{p,\ell}^\pidx{c}(r)\cos(\ell\theta),\\
	R_{p,\ell}^\pidx{c}(r) &= J_\ell(k_{p,\ell}r) + x_{p,\ell}Y_\ell(k_{p,\ell}r),
\end{align}
where $N_{p,\ell}$ is such that $\int_V \mathrm{d}\vb*{r} \lvert\phi_{p,\ell}^\pidx{c}(\vb*{r})\rvert^2 = 1$. Under the simplifying assumption of complete reflection of the electromagnetic field at the boundaries of the ring, the optical wave vector $k_{p,\ell}$ corresponds to the $p$th root of
\begin{equation}
	Y_\ell(k R_i)J_\ell(k R_d) - J_\ell(k R_i)Y_\ell(k R_d) = 0,
\end{equation}
while $x_{p,\ell}$ is given by
\begin{equation}
	x_{p,\ell} = -J_\ell(k_{p,\ell} R_i)/Y_\ell(k_{p,\ell} R_i).
\end{equation}
The optical angular frequency can still be expressed as $\omega_{p,\ell} = k_{p,\ell}c$.

\begin{figure}[tb!]
	\centering
	\includegraphics[width=.8\linewidth]{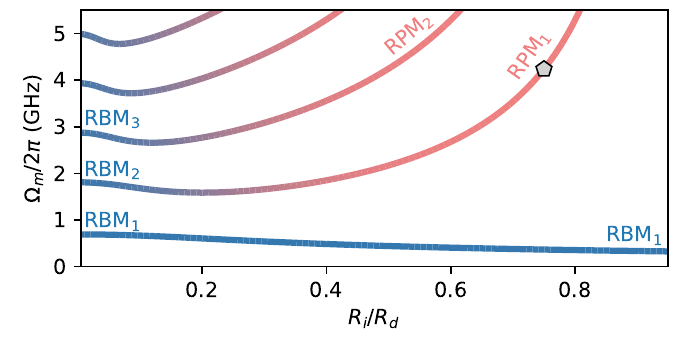}
	\caption{Mechanical spectrum of a ring resonator as the inner radius is increased from $R_i = 0$ (disk) to $R_i \lesssim R_d$ (ring) for an external radius of $R_d = \SI{2}{\um}$ and a thickness of $L_z = \SI{200}{\nm}$. The ring radial pinching mode (RPM) considered in Fig.~\ref{fig:threeExamples} of the main text is indicated with a marker.}
	\label{fig:spectrumRing}
\end{figure}

The presence of a frame supporting the ring has an impact on the mechanical modes. For simplicity, we shall consider a frame composed of $2m$ branches that pin the ring at angular positions separated by $\pi/m$. As a consequence and in contrast with the disk geometry, off-diagonal strain terms $\varepsilon_{r\theta} = \varepsilon_{\theta r}$ no longer vanish and the classical Lagrangian must be adapted as follows:
\begin{equation}
	L_r = \int_V\mathrm{d}\vb*{r}\Bigl\lbrace\frac{1}{2}\rho\dot{u}_r^2 + \frac{1}{2}\rho c_\mathrm{s}^2 u_r \Bigl(\frac{1}{r}\partial_r[r]\partial_r - \frac{1}{r^2} + \frac{\partial_\theta^2}{r^2\Gamma^2}\Bigr)u_r\Bigr\rbrace,
\end{equation}
where $\Gamma = \sqrt{2/(1-\nu)}$ is the effective sound-velocity ratio in the material within the plane-stress conditions. This Lagrangian can be diagonalized by expanding the radial displacement into the following base functions:
\begin{align}
	\phi_{n,m}^\pidx{m}(r,\theta) = \mathcal{N}_{n,m} (&J_{\nu_m}(K_{nm}r)\\
	&+ x_{n,m}Y_{\nu_m}(K_{nm}r))\cos(m\theta),\nonumber
\end{align}
where $\nu_m = \sqrt{1 + (m/\Gamma)^2}$ and $\mathcal{N}_{n,m}$ is a normalization constant such that $\int_V \mathrm{d}\vb*{r} \phi_{n,m}^\pidx{m}(\vb*{r})^2 = V$. The mechanical wave vector $K_{nm}$ is chosen so as to be the $n$th root of
\begin{equation}
	\mathfrak{D}Y_{\nu_m}(K R_i)\mathfrak{D}J_{\nu_m}(K R_d) - \mathfrak{D}J_{\nu_m}(K R_i)\mathfrak{D}Y_{\nu_m}(K R_d) = 0,
\end{equation}
with the differential operator $\mathfrak{D} = \partial_r + \nu/r$; $x_{n,m}$ is given by
\begin{equation}
	x_{n,m} = -\mathfrak{D}J_{\nu_m}(K_{nm} R_i)/\mathfrak{D}Y_{\nu_m}(K_{nm} R_i).
\end{equation}

In the limit where $R_i \rightarrow 0$, one indeed recovers the modes and energies of the disk as derived in the previous appendix, as illustrated in Fig.~\ref{fig:spectrumRing}. At variance, when $R_i \lesssim R_d$, the RBMs of radial order higher than one depart from their original frequencies and acquire a radial-pinching nature, with frequencies scaling as $\sim 1/(R_d - R_i)$. 

\begin{figure}[tb!]
	\centering
	\includegraphics[width=.8\linewidth]{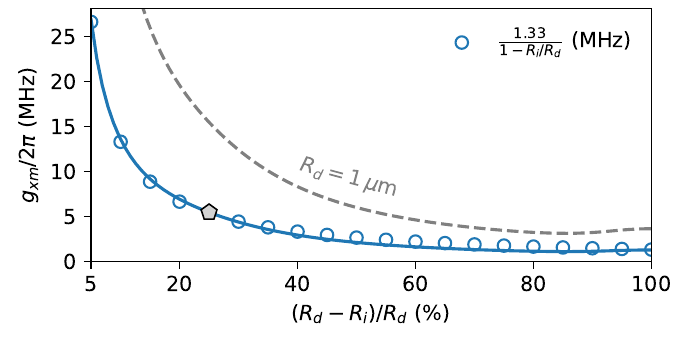}
	\caption{Electromechanical coupling for a ring of external radius $R_d = \SI{2}{\um}$ ($R_d = \SI{1}{\um}$ in dashed lines) and thickness $L_z = \SI{200}{\nm}$ as function of its width $R_d-R_i$. The first radial pinching mode and an excitonic whispering gallery mode of order $(p,\ell) = (1,38)$ are here considered. The value $g_{xm}/2\pi = \SI{5.47}{\MHz}$ considered in Fig.~\ref{fig:threeExamples} of the main text is indicated by a filled marker.}
	\label{fig:gxmRing}
\end{figure}

The above expressions were used in order to evaluate the electromechanical coupling strength of the ring resonator in Fig.~\ref{fig:threeExamples} of the main text. In Fig.~\ref{fig:gxmRing}, we show the value of this parameter and its dependence on the considered WGM indices for a ring of external radius $R_d = \SI{2}{\um}$ and thickness $L_z = \SI{200}{\nm}$, for varying values of its width.

The discussion on the limiting factors for the optical and mechanical decay rate in microrings are analogous to those in microdisks (see Sec.~\hyperref[app:C]{C}). The most important difference is related to the geometry of the resonator: mechanical dissipation trough the tethers supporting the ring modify anchoring losses. We refer to \cite{Anetsberger2008} for details.

\subsection{E - Volume strain in the plane-stress regime\label{app:E}}

In both disk and ring resonators, the considered plane-stress condition $\sigma_{zz} = 0$ causes the strain contribution of the out-of-plane polarization of the mechanical mode to everywhere adapt to the radial strain. Indeed, one has $\varepsilon_{zz} = -\frac{\nu}{1-\nu}(\varepsilon_{rr}+\varepsilon_{\theta\theta})$, and thus finally:
\begin{equation}
	\Sigma(\vb*{r}) = \nabla \cdot \vb*{u}(\vb*{r}) = \frac{1-2\nu}{1-\nu}\frac{1}{r}\partial_r[r] u_r(\vb*{r}).
\end{equation}
The volume strain is thus reduced with respect to the bare in-plane contribution. For $\nu = 0.31$, this reduction is of $45\%$.

\subsection{F - Rabi splitting in a plane microresonator\label{app:F}}

In the considered resonators, quantum-well excitons and TE-polarized
cavity photons are colocalized, and strongly coupled
through the minimal coupling Hamiltonian~\cite{Savona1999,Girlanda1981}:
\begin{equation}
	\hat{H}_{cx} =
	-\frac{e}{m_0}\sum_j\hat{\vb*{A}}(\vb*{r}_j)\cdot\vb*{p}_j +
	\frac{e^2}{2m_0}\sum_j\hat{\vb*{A}}^2(\vb*{r}_j).
\end{equation}
We shall here focus only on its first term, the dominant one provided $ \hat{H}_{cx}$ can be treated as a small perturbation to the bare exciton and photon Hamiltonians. By introducing a set of
excitonic modes with wave functions $\psi_{\lambda'}^\pidx{x}$ and
associated bosonic operators $\hat{d}_{\lambda'}$, one has:
\begin{equation}
	\hat{H}_{cx}^{(\mathrm{I})} =
	ie\sum_{\lambda'}\omega_{\lambda'}^\pidx{x}\Bigl(\bra{0}{\textstyle\sum_j}\hat{\vb*{A}}(\vb*{r}_j)\cdot\vb*{p}_j\ket{\psi_{\lambda'}^\pidx{x}}\hat{d}_{\lambda'}
	+ \mathrm{H.c.}\Bigr),
\end{equation}
where $\omega_{\lambda'}^\pidx{x}$ denotes the angular frequency of the
exciton in the mode $\lambda'$. By further considering a general
circularly polarized vector potential of the form of Eq.~\eqref{eq:30}:
\begin{equation}
	\hat{\vb*{A}}_\sigma(\vb*{r}) =
	\sum_{\lambda}\sqrt{\frac{\hbar}{2\epsilon\omega_\lambda^\pidx{c}}}
	\psi_\lambda^\pidx{c}(\vb*{r})(\hat{a}_\lambda^{\mathstrut} +
	\hat{a}_\lambda^\dagger)\vb*{\epsilon}_\sigma,
\end{equation}
the coupling may be put under the following form:
\begin{equation}
	\hat{H}_{cx}^{(\mathrm{I})} =
	i\sum_{\lambda,\lambda'}C_{\lambda,\lambda'}(\hat{a}_\lambda^{\mathstrut}
	+ \hat{a}_\lambda^\dagger)(\hat{d}_{\lambda'}^{\mathstrut} +
	\hat{d}_{\lambda'}^\dagger),
\end{equation}
with
\begin{align}
	C_{\lambda,\lambda'} &= \sqrt{\frac{\hbar^2 e^2}{2\epsilon
			\omega_\lambda^\pidx{c}}}\omega_{\lambda'}^\pidx{x}\bra{0}{\textstyle\sum_j}\psi_{\lambda}^\pidx{c}(\vb*{r}_j)\vb*{\epsilon}_\sigma\cdot\vb*{r}_j\ket{\psi_{\lambda'}^\pidx{x}}\nonumber\\
	&\simeq
	\int\mathrm{d}\vb*{r}\psi_{\lambda}^\pidx{c}(\vb*{r})\psi_{\lambda'}^\pidx{x}(\vb*{r}_e=\vb*{r},\vb*{r}_h=\vb*{r})[\vb*{\epsilon}_\sigma
	\cdot \vb*{r}_{cv}],
\end{align}
where $\vb*{r}_{cv} = \bra{u_{c,\Gamma}}\vb*{r}\ket{u_{v,\Gamma}}$
denotes the dipole matrix element between the bulk valence- and
conduction-band single-particle Bloch functions. Here, the spatial
variations of both the exciton envelope and the optical mode were
neglected at the scale of the unit cell. By considering normalized wave
functions of the form
\begin{align}
	\psi_{\lambda}^\pidx{c}(\vb*{r}) &=
	F_{\lambda}^\pidx{c}(\vb*{r}_\parallel)f(z),\\
	\psi_{\lambda'}^\pidx{x}(\vb*{r}_e, \vb*{r}_h) &=
	F_{\lambda'}^\pidx{x}(\vb*{R}_\parallel)\phi(\vb*{\rho}_\parallel)\chi_e(z_e)\chi_h(z_h),
\end{align}
finally yields
\begin{equation}
	C_{\lambda,\lambda'} = C_\lambda \mathcal{I}_{\lambda,\lambda'},
\end{equation}
where $\mathcal{I}_{\lambda,\lambda'}$ represents a dimensionless overlap integral, that is diagonal upon assuming a perfect
confinement of the optical field within the resonator:
\begin{equation}
	\mathcal{I}_{\lambda,\lambda'} = \int\mathrm{d}\vb*{r}_\parallel
	F_{\lambda}^\pidx{c}(\vb*{r}_\parallel)F_{\lambda'}^\pidx{x}(\vb*{r}_\parallel)
	\simeq \delta_{\lambda,\lambda'},
\end{equation}
and
\begin{equation}
	C_\lambda = \sqrt{\frac{\hbar^2 e^2}{2\epsilon
			\omega_\lambda^\pidx{c}}}\omega_{\lambda'}^\pidx{x}\phi(\vb*{0})[\vb*{\epsilon}_\sigma
	\cdot \vb*{r}_{cv}]\int\mathrm{d}z\chi_h(z)\chi_e(z)f(z).
\end{equation}
Introducing the oscillator strength
\begin{equation}
	\frac{f_\mathrm{osc}}{S} =
	\frac{2m_0\omega_{\lambda'}^\pidx{x}}{\hbar}\lvert\vb*{\epsilon}_\sigma
	\cdot
	\vb*{r}_{cv}\rvert^2\Bigl\lvert\int\mathrm{d}z\chi_h(z)\chi_e(z)\Bigr\rvert^2\lvert\phi(\vb*{0})\rvert^2,
\end{equation}
and assuming $\omega_\lambda^\pidx{x} \approx \omega_\lambda^\pidx{c}$,
one obtains the expression~\cite{Panzarini1999}
\begin{equation}
	C_\lambda \simeq \hbar\sqrt{\frac{e^2}{2\epsilon_0
			n_{\mathrm{eff}}^2 m_0}\frac{f_\mathrm{osc}/S}{L_\mathrm{eff}}},
\end{equation}
with $n_\mathrm{eff}$ the effective in-plane refractive index and
\begin{equation}
	L_\mathrm{eff} =
	2\left\lvert\frac{\int\mathrm{d}z\chi_h(z)\chi_e(z)}{\int\mathrm{d}z\chi_h(z)\chi_e(z)f(z)}\right\rvert^2
	\simeq 2/f^2(z_\mathrm{QW}).
\end{equation}
This straightforwardly generalizes to the $n$-QW case as
\begin{equation}
	C_{\lambda,n} = \sqrt{n}C_{\lambda},
\end{equation}
with
\begin{equation}
	L_\mathrm{eff}^{-1} = \frac{1}{n}\sum_i L_{\mathrm{eff},i}^{-1}
	\simeq \frac{1}{2n}\sum_i f^2(z_i).
\end{equation}
For the simplest vertical profile, of the form $f(z) =
\sqrt{2/L_z}\cos(\pi z/L_z)$, this simply reads
\begin{equation}
	L_\mathrm{eff}^{-1} = L_z^{-1} \times \frac{1}{n}\sum_i \cos^2(\pi
	z_i/L_z).
\end{equation}

\begin{figure}[tb!]
	\centering
	\includegraphics[width=\linewidth]{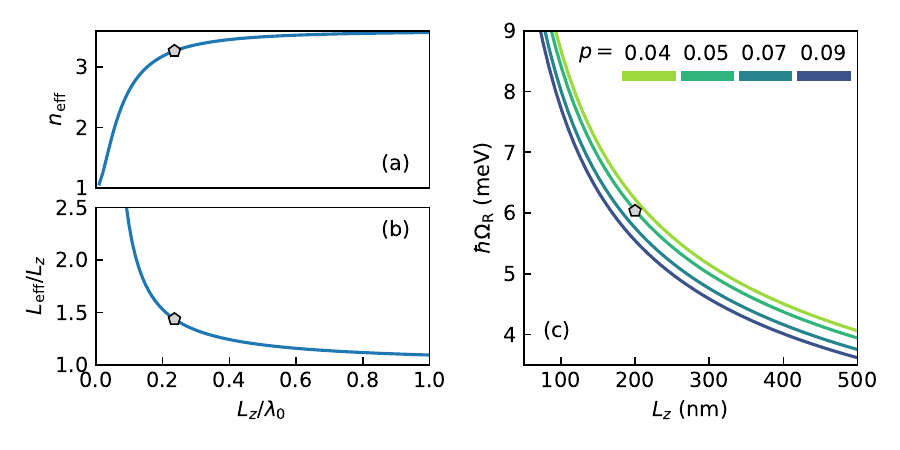}
	\caption{Microdisk/microring: (a) Effective in-plane refractive index and (b) effective length, as approximated by $L_\mathrm{eff} \simeq 2/f^2(z_\mathrm{QW})$, as a function of the vertical confinement in a slab of GaAs. (c) Rabi splitting in a semiconductor microdisk/microring for a $\SI{8}{\nm}$-thick $\mathrm{In}_{p}\mathrm{Ga}_{1-p}\mathrm{As}$ QW. In all three panels, $n_{\mathrm{GaAs}} = 3.60$ ($\lambda_0 \approx \SI{850}{\nm}$). The values of these quantities for the resonators considered in the main text are here denoted by a marker, in particular the single-QW Rabi splitting $\hbar \Omega_\mathrm{R} = \SI{6.04}{\meV}$ ($g_{cx}/2\pi \approx \sqrt{2} \times \SI{0.73}{\THz}$ for $2$ QWs).}
	\label{fig:neffLeffRabi}
\end{figure}

The effective length $L_\mathrm{eff}$ and the Rabi splitting of a plane microresonator are shown in Figs.~\ref{fig:neffLeffRabi}(b) and (c), respectively, as a function of the resonator's thickness $L_z$.

\subsection{G - Optical and vibrational modes: Micropillar\label{app:G}}

A general calculation of the optical modes of a semiconductor micropillar represents a quite involved problem. In the following, we will show that by adopting some controlled approximations, one may derive analytical expressions for the scalar modes of the system. In absence of sources or charges the eigenmode envelopes $\mathbf{E}_n(\mathbf{r})$ of the electric field in the dielectric structure obey the vector-wave equation
\begin{equation}\label{eq:pillarMaxwellEq}
	\nabla \times \nabla \times \mathbf{E}_n(\mathbf{r}) - (\omega_n^2/c^2)\epsilon_r(\mathbf{r})\mathbf{E}_n(\mathbf{r}) = 0,
\end{equation}
where $c$ is the speed of light in vacuum, $\omega_n/2\pi$ is the frequency of each eigenmode and $\epsilon_r$ the spatially-dependent relative permittivity of the dielectric structure. Using the identity $\nabla \times \nabla \times\mathbf{E}= \nabla(\nabla \cdot \mathbf{E}) - \nabla^2 \mathbf{E}$ and the absence of charges $\nabla\cdot (\epsilon_0\epsilon_r \mathbf{E})=0$ one can turn Eq.~\eqref{eq:pillarMaxwellEq} into 
\begin{equation}\label{eq:pillarMaxwellEq2}
	0=\nabla^2 \mathbf{E}_n + \frac{\omega_n ^2}{c^2}\left[\epsilon_r\mathbf{E}_n - \frac{c^2}{\omega^2 _n}\nabla(\epsilon_r^{-1} \nabla \epsilon_r \cdot \mathbf{E}_n) \right].
\end{equation}
Clearly, the right-most term is the only one that couples different components of the electric field. Let us now explicitly consider a pillar of radius $R_p$ vertically defined by a $\lambda/2$ Fabry-Pérot cavity, formed by two mirrored GaAs/AlAs $\lambda/4$ stacks, the $\lambda/2$ spacer being defined by two GaAs layers spliced together at $z=0$. 
Provided the lateral size of the structure is significantly larger than the optical wavelength in the material one can adopt a paraxial approximation. Then, the only relevant components of the electric field in a cylindrical coordinate system are the radial and azimuthal components $E_n ^{r,a}=u_n ^{r,a}(r,\theta)u^z(z)$. As $\epsilon_r$ depends uniquely on the radial and axial coordinates, $\nabla(\epsilon_r^{-1} \nabla \epsilon_r \cdot \mathbf{E}_n)$ contains only $E_r$ and the spatial derivatives of $\log(\epsilon_r)$. These terms, couple the radial and azimuthal components of the electromagnetic field at the pillar boundaries and are responsible for a polarization-dependent fine structure of the eigenmodes~\cite{Whittaker2018}. However, their contribution is generally negligible provided $E_r$ is vanishing at the pillar edge~\cite{Yeh1979}. Given the large refractive index of GaAs, both approximations are justified already for pillars with a diameter of $1$--$\SI{2}{\um}$. At this point, all the equations decouple, and one is left with the problem of determining the envelope $u^{z}$ and two identical scalar Helmholtz equations for $E_n ^{r,a}$. 
In order to find $u^{z}$ one needs to solve the equation $[\nabla^2+\omega^2/c^2 \epsilon_r(z)]u^{z}=0$. Exact solutions can be numerically obtained via the transfer matrix method. In order to derive an approximate expression we can assume that the DBR extends infinitely along $z$ and exploit the mirror symmetry of the problem. For a forward traveling wave, the $z>0$ half cavity is equivalent to a simple DBR, starting with the large refractive index layer. Translational invariance allows finding the envelope $u^{z}$ for $z>0$ by solely determining the propagation constant $\beta_z$ for the DBR. 

This is a simple problem that can be exactly solved using the transfer matrix formalism. If we denote $\mathcal{M}_{1,2}$ the transfer matrices for the two $\lambda/4$ stacks, the transverse wave propagation through one unit cell obeys $\mathcal{M}_{1}\mathcal{M}_{2} = \mathcal{M}_{BZ}$, which is all we need to relate in the first Brillouin zone the energy $E = \hbar c k_z$ of the incident wave to its wavevector in the material. Using the properties of the trace and determinant of $\mathcal{M}_{BZ}$, we can write the dispersion relation
\begin{equation}
	\begin{aligned}
		\cos(d \beta_z)& =\cos(n_1 d_1 k_z)\cos(n_2 d_2 k_z)-\\
		&-\frac{n_{1}^{2}+n_{2}^{2}}{2 n_1 n_2}\sin(n_1 d_1 k_z)\sin(n_2 d_2 k_z),
	\end{aligned}
\end{equation} 
where $d_{1,2}$ and $n_{1,2}$ respectively represent the thicknesses and refractive indices of the layers forming the DBR ($d=d_{1}+d_2$), while the propagation constant $\beta_z=n_{\mathrm{eff}} k_z$ can be expressed product of an effective refractive index of the structure $n_{\mathrm{eff}}$ with the photon wave vector in vacuum $k_z=2\pi/\lambda_0$. Taking $d_{1,2}=\lambda_0/4 n_{1,2}$ and $n_1>n_2$, the first photonic bandgap is centered at $\lambda_0$ where the effective refractive index becomes complex: the real part yields the wave period $\lambda_0/n_{\mathrm{eff}} = \lambda_0 (n_1+n_2)/ 2 n_1 n_2$, whereas the imaginary part corresponds to an exponential decay of the field in the mirror of a typical length $2 \tilde{L}\approx \lambda_0 / 2(n_1-n_2)$~\cite{Savona1999}. 

The same argument holds for the $z<0$ side of the structure, imposing the mirror symmetry and the continuity of the field one finally gets the cavity mode envelope, at least of a normalization constant $\mathcal{N}_z$,
\begin{equation}
	\begin{aligned}
		u^z(z) & =\mathcal{N}_z\, e^{-|z|/2\tilde{L}}\,\sin( n_{\mathrm{eff}} k_0  z),\\
		\mathcal{N}_z&=\sqrt{\frac{1+4 k_0^2 \tilde{L}^2 n_{\mathrm{eff}}^2}{4 k_0^2 \tilde{L}^3 n_{\mathrm{eff}}^2}},
	\end{aligned}
\end{equation}
imposing, as in Appendices~\hyperref[app:C]{C} and \hyperref[app:D]{D}, that $\int^{L_z/2}_{-L_z/2} \mathrm{d}z\, |u^z(z)|^2\approx\int_{\mathbb{R}} \mathrm{d}z\, |u^z(z)|^2=1$. Since we have neglected the finite number of pairs in the structure, we benchmarked these result against the exact envelope obtained via the transfer matrix method. For a cavity formed by two $25.5~\mathrm{GaAs/AlAs}$ stacks ($n_{1}\approx 3.5$ and $n_{2}\approx 2.9$ at $\lambda_0=\SI{0.85}{\um}$ and $4~\mathrm{K}$), we obtain an average deviation below a part per thousand.

The envelopes $u^{r,a}(r,\theta)$ obey identical scalar Helmholtz equations describing the transverse profile of the modes of an infinite waveguide sharing the same cross section as the pillar and characterized by effective refractive index $n_{\mathrm{eff}}$. If the refractive index of the waveguide is large, one can linearize the propagation constant $\beta_z=\beta_0+\delta\beta$ and recast the Helmholtz equation in terms of an equivalent Shr\"{o}dinger equation~\cite{Yeh1979,Marte1997}. The envelopes $u^{r}_n(r,\theta)$ are then solutions to the problem of a particle with an effective mass $m^*_c=E_0 n_{\mathrm{eff}}^2/c^2$ ($E_0=h c/\lambda_0$) in a circular potential well $V(r,\theta)=E_0[n_{\mathrm{eff}}^2-n^2(r)]/2n_{\mathrm{eff}}^2$, where $n(r)$ denotes the refractive index profile along the radial coordinate. For a large refractive index contrast, the potential barrier for the guided modes is $\Delta E\sim E_0/2$. At the same time, within the paraxial approximation $(\delta\beta_n/\beta_0)^2=(\delta E_n/E_0)^2\ll 1$: one can thus fairly approximate the effective potential with an infinite circular well. This is a known problem: the radial part of the Hamiltonian can be diagonalized in the basis of Bessel functions of the first kind  $J_{n}(r)$ while the angular component of the envelope remains associated to the angular momentum $e^{i\ell\theta}$. Specializing to the fundamental mode of the micropillar,  $\ell=0$ and 
\begin{equation}\label{eq:radialEnvOptMode_Pillar}
	\begin{aligned}
		u^r(r) & =\mathcal{N}_r J_0 ( K_1 r ),\\
		\mathcal{N}^{-1}_r & = \sqrt{\pi} J_{1}(\alpha_{01})R_p.
	\end{aligned}
\end{equation}  
Here $K_1=\alpha_{01}/R_p$, with $\alpha_{01}$ the first zero of $J_0(r)$, and $R_p$ is the pillar radius. We impose the normalization condition $2\pi\int |u^r(r)|^2 r\,\mathrm{d}r=1$.

Finally, we can exploit the fact that phonons are perfectly co-localized with photons in GaAs/AlAs microresonators due to the nearly identical optical and acoustic impedance mismatch of the two materials~\cite{Fainstein2013} to derive the strain field envelope. Besides narrow periodic windows in $R_p$ where the Poisson ratio in the material couples the longitudinal and radial components of the displacement field at the pillar boundary, the elastic energy of the fundamental mode is almost entirely stored in the longitudinal component~\cite{Anguiano2018}. Then, by imposing stress-free conditions at the pillar boundary, allows writing the strain field associated to the fundamental modes as
\begin{equation}\label{eq:strainField_Pillar}
	\Sigma(r,z)  = \mathcal{N}_\Sigma \, u^{r}(r) \partial_z u^{z}(z),
\end{equation}
where $\mathcal{N}_{z}^{-1}\mathcal{N}_{r}^{-1}\mathcal{N}_\Sigma= \beta^{-1}_{z} \cos^{-1}[(1+4\beta_z^2\tilde{L}^2)^{-1/2}]\approx e^{\Delta n / 2 n_{\mathrm{eff}}}$ ensures a unit normalization of the displacement field amplitude at the reduction point of the mechanical mode~\cite{Hauer2013}. Finally, we can find an approximate expression for the electromechanical coupling. Inserting Eq.~\eqref{eq:radialEnvOptMode_Pillar}--\eqref{eq:strainField_Pillar} in Eq.~\eqref{eq:approxGxm} of the main text and using the identity $\int_0^{R_p} \mathrm{d}R\, 2\pi R J_0 ( K_n r )^3 = \beta_0 J_{1}(\alpha_{01})R_p^2$, with $\beta_0\approx 1.18$ a numerically-determined constant, we get 
\begin{equation}
	G_{xm}  = (a_h-a_e)k_m\, \eta_{S}\,\mathcal{I}_{g} ,
\end{equation}    
where $k_m=2\pi n_{\mathrm{eff}}/\lambda_0$ is the phonon wave vector, $\eta_{S}=\Sigma(0,z_{QW})/|\Sigma_{max}|$, $|\Sigma_{max}|$ is the strain field maximum, and $\mathcal{I}_{g}$ represents the geometric overlap integral associated to the pillar electromechanical coupling:
\begin{equation}
	\mathcal{I}_{g}  = \frac{\beta_0}{ \pi J_{1}(\alpha_{01})}\exp\left(\frac{\Delta n}{2n_{\mathrm{eff}}}\right).
\end{equation}
As an example, we consider a cavity formed by two $25.5~\mathrm{GaAs/AlAs}$ stacks with central wavelength $\lambda_0=\SI{0.85}{\um}$ and obtain  $G_{xm}\eta_{S}^{-1}\approx 2\pi \times \SI{44.3}{\THz/\nm}$. As the Bohr radius satisfies $n_{eff} a_B/\lambda_0\ll1$, this approximate result departs less than a percent from the one obtained by a direct numerical evaluation of Eq.~\eqref{eq:exactGxm}.

The exciton frequency shift induced by a single phonon $g_{xm}$ can be determined upon finding $x_\mathrm{ZPF}=\sqrt{\hbar/2 \tilde{m} \Omega_m}$. The angular frequency for the fundamental longitudinal mode is given by $\Omega_m^2=\Omega_{m,0}^2+  \alpha_{01}^2 (\tilde{v}_S/2 R_p)^2$ where $R_p$ is the pillar radius $\Omega_{m,0}\approx 2\pi \times \SI{19.5}{\GHz}$ is the cutoff frequency of the planar cavity and $\tilde{v}_S=\int \mathrm{d}z \, |u^{z}(z)|^2 v_{S}(z) \approx \SI{5.27}{\um\per\ps}$ is the effective speed of sound in the multilayer structure~\cite{Anguiano2018}. Using the radial homogeneity of the density in the resonator, one can determine the effective mass using $\tilde{m}= \pi J_1(\alpha_{01})^2 R_p^2 \,\tilde{\rho}$ with $\tilde{\rho}=\int \mathrm{d}z \, |u^{z}(z)|^2  \rho(z)$. Notice that the effective mass of the mode scales with $\tilde{L}\sim \lambda_0 \sim k_m^{-1}$ by virtue of the photon-phonon co-localization along the pillar axis. This cancels the $\Omega$ dependence of $x_{\mathrm{ZPF}}$, and thus $g_{xm}\sim k_m$. If we consider a pillar with $\SI{2.6}{\um}$ diameter, one gets $\Omega_m=2\pi \times \SI{19.6}{\GHz}$, $\tilde{m}=\SI{0.7}{\pg}$ and $x_\mathrm{ZPF}=\SI{0.8}{\femto\meter}$. The optomechanical coupling $g_{cm}$ for this kind of structure was already calculated numerically in \cite{Anguiano2018} obtaining the value $G_{cm}\approx 2\pi \times \SI{0.43}{\THz/\nm}$. Finally, the light-matter coupling $g_{cx}$ for a single QW reads \cite{Panzarini1999,Savona1999}:
\begin{equation}
	g_{cx} \approx \tilde{g}_{cx} \eta_E =2\pi\sqrt{\frac{2 c \Gamma_x}{\hbar n_{\mathrm{eff}}\tilde{L}}}\eta_E,
\end{equation}
where $\Gamma_x$ is the exciton half-linewidth calculated in Appendix~\hyperref[app:B]{B} (cf.~Fig.~\ref{fig:ShallowQWs_Excitons}), $L_{\mathrm{eff}}=2\tilde{L}+L_{sp}$ is the effective length of the cavity ($L_{sp}=\lambda_0/2n_{\mathrm{eff}}$ is the spacer optical thickness) and $\eta_E=|E(z_{QW})/E_{\mathrm{max}}|$ is the reduced amplitude of the electric field in the QW plane. Considering a set of four QWs displaced by $\Delta z_j=\pm(15,39)~\mathrm{nm}$ from the node of the cavity field, as for the pillar described in the main text, one has $\eta_{E,j}=(0.35,0.80)$, and the effective coupling for the bright polariton state becomes $g_{cx}=\tilde{g}_{cx}(\sum_j \eta_{E,j}^2)^{1/2}\approx 2\pi\times 0.53~\mathrm{THz}$. The effective value of $G_{xm}$ when considering a structure embedding multiple QWs (each in the strong exciton-photon coupling regime) can be calculated as $G_{xm}=(a_h-a_e) k_m \mathcal{I}_g (\sum_{j}\eta_{E,j}^2\eta_{\Sigma,j})/(\sum_{j}\eta_{E,j}^2)$. For the above considered multiple QW arrangement, we have $\eta_{\Sigma}=(0.93,0.62)$ yielding $G_{xm}\approx 2\pi\times 30~\mathrm{THz/nm}$.\\

The optical and mechanical quality factor play an important role in determining the optomechanical cooperativity discussed in the main text: we briefly summarize the factors determining the two below. Concerning the optical decay rate, we have three main factors determining the intrinsic value of $\kappa_c$, namely radiative decay, residual absorption and coupling to leaky modes. Radiative losses are determined by the number of DBR pairs. Due to the finite optical penetration depth ($\tilde{L}$), thicker mirrors have a better reflectivity, as the overlap of the cavity mode tails with free-space modes is suppressed. For a wave at normal incidence, the reflectivity of a DBR is $R_c=|r(\omega_0)|^2=1-4(n_{\mathrm{out}}/n_{\mathrm{in}})(n_2/n_1)^{2N_{\mathrm{DBR}}}$, where $n_{\mathrm{in}}$ and $n_{\mathrm{out}}$ denote the refractive indices of the medium before and after the DBR \cite{Savona1999}. In order to have two mirrors with a nearly identical reflectivity, the DBR facing the GaAs substrate must have 3 pairs more since $1/n_{1}\approx (n2/n1)^6$. Assuming $(1-R_c)/R_c\ll 1$ and a symmetric cavity configuration, the radiative decay rate for a $\lambda/2$ cavity is
\begin{equation}\label{eq:DBRradiativelosses}
    \kappa_{c,r}/2\pi\approx \left(\frac{c}{\lambda_0}\right)\left(\frac{n_1^2-n_2^2}{n_1 n_2}\right)\frac{1-\sqrt{R_c}}{4\sqrt{R_c}}
\end{equation}
and, in principle, it can be made arbitrarily small by increasing the DBR pairs number ($N_{\mathrm{DBR}}$). In practice, residual absorption in GaAs imposes a fundamental limit to the optical quality factor. Recent experiments \cite{CarlonZambon2020th} (Chapters 5 and 6) report a typical absorption rate of $\kappa_{c,a}/2\pi\sim 2 ~\mathrm{GHz}$ for modes around $850~\mathrm{nm}$ at 4K. Finally, due to the finite angular acceptance of the photonic bandgap created by the DBRs, an extreme lateral confinement of the optical field in micropillars eventually couples the cavity mode to leaky modes with a large transverse wave-vector. The effect becomes dramatic for pillars presenting a radius comparable with the optical wavelength in the material $R_p \sim \lambda_0/n_{\mathrm{eff}}$, i.e. when the paraxial approximation breaks down. This and other effects, as sidewall inclination, are investigated in detail in \cite{Karl2009}. This latter effect is the only intrinsic size-dependent contribution to the optical quality factor and for the moderate values of $Q_c \sim \num{50e3}$ described in the main text and for pillar radii $d\gtrsim 0.6~\mathrm{\mu m}$ it should not play a significant role \cite{Karl2009}. Therefore, for simplicity in the main text we consider $\kappa_c\approx \kappa_{c,r}+\kappa_{c,a}$, independent of the pillar radius.\\

Due to the co-localization of photons and phonons in GaAs/AlAs heterostructures \cite{Fainstein2013}, mechanical losses associated to coupling of the acoustic mode to the substrate obey an expression nearly identical to Eq.~\eqref{eq:DBRradiativelosses} upon replacing the optical DBR reflectivity with the acoustic DBR reflectivity $R_a=1-4(\rho_{\mathrm{out}} v_{\mathrm{out}}/\rho_{\mathrm{in}} v_{\mathrm{in}})(\rho_2 v_2/\rho_1 v_1)^{2N_{\mathrm{DBR}}}$, where $\rho_i$ and $v_i$ denote the mass density and speed of sound in the material. Notice that one of the top mirror faces vacuum, and will thus have a unit reflectivity ($\rho_{\mathrm{out}}=0$), thus $Q_m\approx Q_c$. Several parasitic effects can degrade the mechanical quality factor: for GaAs resonators, surface roughness and microscopic defects in the material acting as relaxing two-level systems (TLS) are the dominant ones \cite{Hamoumi2018,Anguiano2018}. Nevertheless, surface passivation techniques and operation in a cryogenic environment can be exploited to effectively suppress these effects \cite{Hamoumi2018}.

\subsection{H - Input-output theory and stability analysis\label{app:H}}

In the following, we consider coherent driving of the optical mode with a narrow-band oscillator at a frequency $\omega$. If the driving tone is nearly resonant with the lower polariton resonance $(\omega_0-\omega_l)/2 g_{cx}\ll 1$, as $g_{cx}/\kappa_{c}\gg 1 $ in the strong-coupling regime, we can neglect the upper polariton dynamics. For a two-sided symmetric cavity, the finite reflectivity of the mirrors opens two ports for vacuum fluctuations to enter the system over a bandwidth determined by the optical mode linewidth $\hbar\kappa_c$. Additionally, non-radiative exciton recombination at a rate $\kappa_x$ can also participate to the dissipation of excitations in the system, resulting in an overall lower polariton decay rate $\kappa_l=(\kappa_{c} \cos^2\theta_{cx} + \kappa_{x} \sin^2\theta_{cx})=(\kappa_r+\kappa_{nr})$, sum of a radiative and of a non-radiative component. Notice that we are neglecting the terms stemming from the inhomogeneous broadening of the matter transition~\cite{Diniz2011}, as we require in order to exert a coherent control over the mechanical motion that $\Gamma_{inh}/\Gamma_m \ll 1$ and $\Gamma_{m}/\kappa_l \ll 1$. Similarly, the mechanical dissipation rate $\Gamma_m$ defines the coupling strength with the phonon thermal bath. 

\begin{figure}[tb]
	\centering
	\includegraphics[trim=0cm 0cm 0cm 0cm, width=86mm]{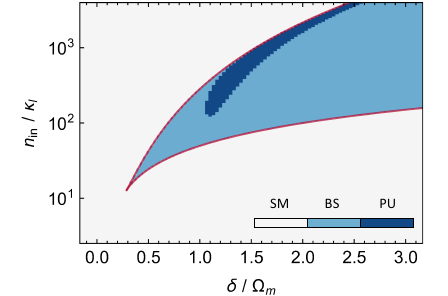}
	\caption{Steady-state behaviour: Stability diagram obtained by inspection of the stability matrix eigenvalues for each available fixed point as a function of the laser detuning $\delta$ and input photon rate $n_{in}$. The character of the steady-state is color coded: SM, BS and PU denote respectively single mode, bistable and parametrically unstable regions in parameter space. The solid purple line corresponds to the analytical bound for the BS region derived upon inserting Eq.~\eqref{eq:KerrBistBound} in \eqref{eq:KerrSSeq}. Here we consider repulsive Kerr interaction; all the parameters are chosen as in Fig.~\ref{fig:NonlinearSBcool} in the main text.}
	\label{fig:StabDiagram}
\end{figure}

Starting from the the Hamiltonian describing the coupled lower-polariton and mechanical modes one can derive the following input-output relations ($\hbar=1$):
\begin{equation}\label{eq:IOequations}
	\begin{aligned}
		i\dot{\hat{l}}& =(-\delta_l - i \kappa_l/2 +\chi_l (\hat{n}_l-1)  - G_{lm} \hat{x}) \hat{l} + i \hat{F}_l + \hat{\xi}_l,\\
		\dot{\hat{p}}&=-\tilde{m} \Omega_m^2 \hat{x} -\Gamma_m \hat{p} + G_{lx} \hat{n}_l +\hat{\xi}_m,\\
		\dot{\hat{x}}&=\hat{p}/\tilde{m},
	\end{aligned}	
\end{equation}
where $\delta_l=(\omega_0-\omega_l)$ defines the laser detuning; $\kappa_l$ and $\chi_l$ define the lower polariton linewidth and Kerr-shift; $G_{lm}$ is the effective optomechanical coupling strength; $\hat{F}_{l}=\sqrt{\kappa_r/2}\,\hat{a}_{in}$, $\hat{a}_{in}$ representing the driving laser field satisfying $\langle \hat{a}^{\dag}_{in}\hat{a}_{in}\rangle=\hbar\omega_0 P$ with $P$  the incident photon rate; $\Omega_m$ and $\Gamma_m$ are the mechanical mode frequency and damping; $\hat{\xi}_{l,m}$ are noise operators satisfying 
\begin{equation}
	\langle \hat{\xi}_l (t) \hat{\xi}^{\dag}_l (t') \rangle  = \kappa_l \delta(t-t'),
\end{equation}
with all the other correlators in $\hat{\xi}_l$ being zero, as the number of thermal excitations at optical frequencies is negligible, and
\begin{equation}
	\langle \hat{\xi}_m (t) \hat{\xi}_m (t') \rangle  = 2\tilde{m} \Gamma_m \int \frac{\mathrm{d}\omega}{2\pi}\, \omega e^{i\omega (t-t')} (\overline{n}_{th}(\omega)+1),
\end{equation}
with $\overline{n}_{th}(\omega)=(e^{\hbar\omega/k_B T}-1)^{-1}$, $k_B$ the Boltzmann constant and $T$ the thermal bath temperature. Hereafter, to ease the notation, we drop all the unnecessary subscripts, keeping only those to distinguish between the noise operators. In order to calculate the steady-state expectation values for the polariton field $\alpha =\langle l \rangle$ and for the mechanical displacement $ q =\langle \hat{x} \rangle$, we can take the expectation of both the terms in Eqs.~\eqref{eq:IOequations}. Using a mean-field approximation for higher than second order correlators and recalling that the noise terms have zero average, one finds:
\begin{equation}\label{eq:MFIOequations}
	\begin{aligned}
		0& =(-\delta - i \kappa/2 +\chi \tilde{n}_{\alpha}  - G \tilde{x})\tilde{\alpha} + i \tilde{F}, \\
		0&=-\tilde{m} \Omega^2 \tilde{q}  + G \tilde{n}_{\alpha},
	\end{aligned}	
\end{equation} 
where we have denoted the steady state solutions satisfying $\dot{\alpha}=\dot{q}=0$ with an over-tilde and $\tilde{n}_{\alpha}=|\tilde{\alpha}|^2$. The second equation is telling us that on average each polariton exerts a deformation $\tilde{q}=G \tilde{n}_{\alpha}/\tilde{m} \Omega^2$: once inserted in the first equation and using the relation $G^2 \equiv 2 \tilde{m} \Omega g^2$, this in turn results into Kerr-like red-shift $2 g^2 \tilde{n}_{\alpha} /\Omega$ of the polariton resonance. It is then convenient to define an effective Kerr coefficient $\tilde{\chi}=(\chi-2 g^2/\Omega)$. Upon introducing $\tilde{\chi}$, one can multiply the first equation by its conjugate to obtain a cubic algebraic equation in the steady-state polariton number $\tilde{n}_{\alpha}$ and input photon rate $n_{in}$:
\begin{equation}\label{eq:KerrSSeq}
	n_{in}=\frac{2}{\kappa_r}[(\delta-\tilde{\chi} \tilde{n}_{\alpha})^2+\kappa^2/4]\tilde{n}_{\alpha}.
\end{equation}
The above equation admits multiple roots in a region of parameter space bounded by the condition that the solutions $\tilde{n}_{\pm}$ of $\partial n_{in} / \partial \tilde{n}_{\alpha} = 0$ are non degenerate. Since
\begin{equation}\label{eq:KerrBistBound}
	\tilde{n}_{\pm}=\frac{2\delta}{3\tilde{\chi}}\pm \frac{1}{6\tilde{\chi}}\sqrt{4\delta^2-3\kappa^2},
\end{equation}
this is the case when $\delta/\kappa > \sqrt{3}/2$. For a given input-photon rate, three possible solutions are available in this region: two are dynamically stable solutions with $\tilde{n}_{\alpha}\gtrless \tilde{n}_{\pm}$ and one is (single-mode) unstable as $\partial n_{in} / \partial \tilde{n}_{\alpha} < 0$. The region parametrized by the condition $\tilde{n}_{-}<\tilde{n}_{\alpha}<\tilde{n}_{+}$ will then be systematically excluded when studying the linearized Hamiltonian as it does not support any stable attractor, and corresponds to the shaded gray areas in Fig.~\ref{fig:NonlinearSBcool} in the main text. 
Notice that a bare Kerr resonator under near-resonant monochromatic driving does not support parametric instabilities. This is not necessarily the case here owing to the parametric coupling to the mechanical mode, which can eventually trigger mechanical self-oscillations~\cite{Kippenberg2005,Aspelmeyer2014}. We will here see how to derive analytical bounds for the region where such parametric instabilities arise. For the sake of completeness, in the following, we alternatively derive the Bogolyubov stability matrix for the coupled system, whose spectrum encodes the temporal evolution of small perturbations driven by the random forces $\hat{\xi}(t)$. The stability matrix can be calculated from the dynamical equations for the expectation values of the polariton ($\tilde{\alpha}$) and phonon ($\tilde{\beta}$) ladder operators, with $\tilde{\beta}$ canonically conjugated to $\tilde{x}$ and $\tilde{p}$ in Eq.~\eqref{eq:MFIOequations}. According to the Grobman-Hartman theorem, the dynamical stability of a fixed point $(\tilde{\alpha},\tilde{\beta})$ with respect to small perturbations is encoded in the eigenvalues of the Jacobian of the vector field determining the dynamics of the system $(\dot{\alpha},\dot{\beta})^T=\mathcal{V}(\alpha,\beta)$. As $\mathcal{V}$ here is analytic but not holomorphic, the Jacobian $J_{\mathcal{V}}$ needs to be calculated using the Wirtinger calculus conventions. The eigenvalues $\omega$ of the stability matrix are the roots of the secular equation $|J_{\mathcal{V}}-\omega \mathbb{1}|=0$ i.e. 
\begin{equation}\label{eq:StabMat}
	0=\begin{vmatrix}
		\Xi_p^{-1} (\omega) &  \chi\tilde{\alpha}^2 & - g \tilde{\alpha} & - g \tilde{\alpha} \\
		-\chi(\tilde{\alpha}^*)^2 & -\Xi_p^{-1} (-\omega)^* &  g \tilde{\alpha}^* &  g \tilde{\alpha}^* \\
		-g \tilde{\alpha}^* & -g \tilde{\alpha} & \Xi_c^{-1} (\omega) & 0 \\
		g \tilde{\alpha}^* & g \tilde{\alpha}  & 0 & -\Xi_c^{-1} (-\omega)^*\\
	\end{vmatrix},
\end{equation}
where we have introduced the quantities $\Xi_p^{-1} (\omega)=(-\tilde{\delta}-i\kappa/2-\omega)$, with $\tilde{\delta}=(\delta-2\tilde{\chi}\tilde{n}_{\alpha})$ and $\Xi_c^{-1} (\omega)=(\Omega-i\Gamma/2-\omega)$, physically representing the bare oscillators susceptibilities. If at least one among the eigenvalues presents an imaginary part larger than zero the system is dynamically unstable, with the real part dictating the single mode ($\mathrm{Re}(\omega)=0$) or parametric ($|\mathrm{Re}(\omega)|>0$) nature of the instability. In Fig.~\ref{fig:StabDiagram} we explore the stability and character of the steady-state of the coupled system by solving Eq.~\eqref{eq:KerrSSeq} and \eqref{eq:StabMat} as a function of the laser detuning and input photon rate. The other parameters are chosen as in Fig.~\ref{fig:NonlinearSBcool} in the main text. 

We can see that the bistable region in parameter space is fairly described by the analytical expression (solid purple line) that we can obtain for the bare Kerr resonator subject to an effective nonlinearity $\tilde{\chi}$. The optomechanical coupling does not sizably perturb the boundaries of this region provided $2g^2/\Omega\chi \ll 1$, which is generally the case in our system. Indeed, on one hand we have $\chi\approx g_{xx} \sin^4\theta_{cx} / \mathcal{A}$, where $\mathcal{A}=\pi J_{1}(\alpha_{10})^2R_p^2$ represents the transverse modal area. On the other one, if we neglect the optomechanical interaction in front of the dominant electromechanical term, we have that $2 g^2 \approx \hbar G^2 \sin^4\theta_{cx}  / \Omega \mathcal{A} \tilde{\rho}$. Therefore the ratio of the two Kerr-shifts does not depend on the exciton fraction nor on the pillar radius, and we get $2 g^2/\Omega\chi \approx \hbar G_{xm}^2/ \tilde{\rho}\,\Omega^2 g_{xx}\sim 10^{-4}$ for an exciton-exciton interaction constant  $g_{xx}\sim \SI{20}{\micro\eV\micro\meter\squared}$~\cite{Munoz-Matutano2019,Delteil2019}.

\subsection{I - Linearized equations and squeezing transformation\label{app:I}}

Here we derive the linearized equations governing the evolution of small fluctuations about the steady-state of the system. Rewriting the operators in terms of their expectation value plus fluctuations $\hat{l}=\tilde{\alpha}+\delta\hat{\alpha}$ and $\hat{q}=\tilde{q}+\delta\hat{q}$ in the hamiltonian and neglecting higher than second order terms in the fluctuations, that is a Gaussian truncation in the correlation hierarchy, we get the Langevin equations:
\begin{equation}\label{eq:linearizedEOMs}
	\begin{aligned}
		& i\dot{\delta\hat{z}} =(-\tilde{\delta}-i\kappa/2)\delta\hat{z} + \chi \tilde{n}\delta\hat{z}^{\dag} - G \tilde{n} \delta\hat{q} + \tilde{\alpha}^*\hat{\xi}_l(t),  \\
		&\tilde{m}(\ddot{\delta\hat{q}}+\Omega^2\delta\hat{q} +\Gamma \dot{\delta\hat{q}})= G (\delta{\hat{z}}+\delta{\hat{z}}^{\dag}) + \hat{\xi}_m(t),
	\end{aligned}	
\end{equation}
where we have multiplied both sides of the first equation by $\tilde{\alpha}^*$ and introduced the rescaled variable $\delta\hat{z}=\tilde{\alpha}^*\delta\hat{\alpha}$ in order to eliminate unnecessary global phases. Furthermore, to ease the notation we hereafter drop the operator symbols, tacitly assuming $[\delta z,\delta z^{\dag}]=\tilde{n}$. These equations faithfully describe the fluctuations dynamics provided non-classical correlations are vanishing in the steady-state, that is single-polaritons nonlinearities need to be weak $\chi/\kappa\ll 1$~\cite{Carusotto2013,Bonifacio1978,Drummond1980}. Due to the squeezing term $\chi\tilde{n}\delta\hat{z}^{\dag}$ in Eq.~\eqref{eq:linearizedEOMs} the dynamics of $\delta z$ and $\delta z^{\dag}$ is coupled. We can write it compactly as $\dot{\vec{z}}=\mathcal{M}\vec{z}-G\tilde{n}\vec{e}\, \delta q + \vec{\xi}$, where $i\vec{z}=(\delta z, \delta z^{\dag})^{T}$, $\vec{e}=(1,-1)^{T}$, $\vec{\xi}=(\tilde{\alpha}^*\xi_l,-\tilde{\alpha}\xi^{\dag}_l)^T$ and 
\begin{equation}
	\mathcal{M}=\begin{pmatrix}
		-\tilde{\delta}-i\kappa/2 & \chi \tilde{n}\\
		-\chi \tilde{n} & \tilde{\delta}-i\kappa/2\\
	\end{pmatrix}.
\end{equation}

Following the approach outlined in refs.~\cite{Clark2017,Asjad2019} 	one can then find a canonical transformation $U_s$ that makes $\mathcal{M}$ diagonal for some \textit{squeezed} displacement operators defined by $\vec{s}=U_s\,\vec{z}$. In order to find $U_s$, we can start by parametrizing a transformation in $\mathrm{SL}(1,1)$ via
\begin{equation}
	U_s(r,\theta_s)=\begin{pmatrix}
		\cosh(r) & e^{i\theta_s}\sinh(r)\\
		e^{-i\theta_s}\sinh(r) & \cosh(r)
	\end{pmatrix},
\end{equation}
thus defining the squeezing parameter $r$ and angle $\theta_s$ (a global phase has been gauged out). Imposing $\mathcal{M}=U_s^{-1}\mathcal{M}_s\,U_s$ for a target $\mathcal{M}_s=\mathrm{diag}(-\delta_s-i\kappa/2,+\delta_s-i\kappa/2)$, as $\chi\tilde{n}\in{\rm I\!R}$, the set of equations yields $\theta_s=0$ (for $\theta_s\in[-\pi/2,\pi/2]$) and 
\begin{equation}\label{eq:SqueezingParameter}
	r=\frac{1}{2}\mathrm{arctanh}\left(\frac{-\chi\tilde{n}}{\tilde{\delta}}\right).
\end{equation}
The fact that $\theta_s=0$ implies that $\delta z=s \cosh(r) - s^{\dag}\sinh(r)$ and using $\mathcal{M}_s=U_s(r,0) \mathcal{M} U^{-1}_s(r,0) $ we find
\begin{equation}\label{eq:SqueezedFieldDetuning}
	\delta_s=\tilde{\delta}\cosh(2r)+\chi \tilde{n}\sinh(2r).
\end{equation}
Concerning the optomechanical coupling one has that $(G\tilde{n})\vec{e}=(G_s\tilde{n})U_s^{-1}\vec{e}$ and again, since $\theta_s=0$ we get the effective coupling coefficient:
\begin{equation}\label{eq:EffOptomechCoupling}
	G_s=G e^{-r}.
\end{equation} 
If we define the correlation matrix $\mathcal{C}$ for the input noise operators via $\langle \vec{\xi}(t)\vec{\xi}(t')^{T}\rangle=\mathcal{C}\,\delta(t-t')$ (recall that $\langle \xi_l(t)\xi_l^{\dag}(t')\rangle = \kappa \delta(t-t')$ while all the other correlators are zero), we have that $\vec{\xi}_s=U_s\vec{\xi}$ and the correlation matrix $\mathcal{C}_s$ for the transformed input noise operators reads $\mathcal{C}_s=U_s \mathcal{C} U_s^{-1}$, i.e.
\begin{equation}\label{eq:CorrMatrix_SqueezedBath}
	\mathcal{C}_s= \kappa \tilde{n} \begin{pmatrix}
		m_s & n_s+1\\
		n_s & m_s
	\end{pmatrix},
\end{equation}
where $n_s=\sinh^2(r)$ defines an effective population of the squeezed optical bath and $m_s=\sqrt{n_s(n_s+1)}$. The equations of motion for the squeezed displacement operators finally become
\begin{equation}\label{eq:squeezedLinearizedEOMs}
	\begin{aligned}
		& i\dot{s} =(-\delta_s-i\kappa/2)s - G_s \tilde{n} \delta q + \xi_s(t), \\
		&\tilde{m}(\ddot{\delta q}+\Omega^2\delta q +\Gamma \dot{\delta q})= G_s (s+s^{\dag}) + \hat{\xi}_m(t),
	\end{aligned}
\end{equation}
As it becomes clear looking at the above expressions, the main advantage of performing the squeezing transformation is that we can recast Eqs.~\eqref{eq:squeezedLinearizedEOMs} into the standard problem of an harmonic optical resonator coupled to a mechanical mode, upon introducing some effective parameters $(\delta_s,G_s,n_s)$ and a squeezed optical bath described by the input noise operators $\xi_s$.

\subsection{J - Mechanical displacement spectrum\label{app:J}}

In order to evaluate the mechanical displacement spectrum it is convenient to move to the frequency domain. We adopt the convention $s_{\omega}=\int_{\mathbb{R}} \mathrm{d}t \, e^{i\omega t} s_t$ to transform Eqs.~\eqref{eq:squeezedLinearizedEOMs} into
\begin{equation}\label{eq:linearizedEOMs_freqDom}
	\begin{aligned}
		& 0  =\Xi_s(\omega)^{-1} s_{\omega} - G_s \tilde{n} \delta q_{\omega} + \xi_s(\omega), \\
		&\Xi_m(\omega)^{-1}\delta q_{\omega}= G_s (s_{\omega}+s^{\dag}_{\omega}) + \hat{\xi}_m(t),
	\end{aligned}	
\end{equation}
where $\Xi_s(\omega)^{-1}=(-\delta_s-i\kappa/2-\omega)$ and $\Xi_{m}(\omega)^{-1}=\tilde{m}(\Omega^2-\omega^2-i\omega\Gamma)$. Using the relation $s^{\dag}_{\omega}=(s_{-\omega})^{\dag}$, some simple algebra yields the spectrum of the mechanical displacement:
\begin{equation}\label{eq:dispSpectrum}
	\delta q_{\omega} = \Xi_\mathrm{eff}(\omega)[\xi_m(\omega) - G_s\xi_{om}(\omega)],
\end{equation}
where $\Xi_\mathrm{eff}(\omega)$ represents the effective mechanical susceptibility modified by the optomechanical interaction
\begin{equation}\label{eq:EffMechsusceptibility}
	\Xi_\mathrm{eff}(\omega)^{-1}=\Xi_m(\omega)^{-1}-G_s^2\tilde{n}\,\left[\Xi_s(\omega)+\Xi_s^{*}(-\omega)\right],
\end{equation}
and $\xi_{om}(\omega)$ is the spectrum of radiation-pressure induced fluctuations actuating the mechanical mode
\begin{equation}\label{eq:OMfluctSpcetrum}
	\xi_{om}(\omega)=\Xi_s(\omega)\xi_s(\omega)+\Xi_s^{*}(-\omega)\xi_s^{\dag}(\omega).
\end{equation}
From the optomechanical self-energy $\Sigma_s(\omega)=\Xi_\mathrm{eff}(\omega)^{-1}-\Xi_{m}(\omega)^{-1}$ we can derive the frequency-dependent corrections to the mechanical frequency $\delta\Omega_{\omega}=\mathrm{Re}[\Sigma_s(\omega)]/2\omega_m$ and optomechanical damping rate $\delta\Gamma_{\omega}=-\mathrm{Im}[(\Sigma_s(\omega)]/\omega_m$:
\begin{equation}\label{eq:OMfreqShift}
	\delta\Omega_{\omega}= \frac{\tilde{g}^2_s\Omega}{\omega}\left[ \frac{\delta_s+\omega}{(\delta_s+\omega)^2+\kappa^2/4} + \frac{\delta_s-\omega}{(\delta_s-\omega)^2+\kappa^2/4}\right],
\end{equation}
\begin{equation}\label{eq:OMlinewidth}	
	\delta\Gamma_{\omega}=\frac{\tilde{g}^2_s\Omega}{\omega}\left[ \frac{\kappa}{(\delta_s+\omega)^2+\kappa^2/4} - \frac{\kappa}{(\delta_s-\omega)^2+\kappa^2/4}\right],
\end{equation}
where we introduced the quantity $\tilde{g}^2_s=\tilde{n}G^2_s/2\tilde{m}\Omega$. In absence of optical nonlinearities the above expressions reduce to the well-known relations for the optical spring effect and optomechanical damping~\cite{Aspelmeyer2014} as the squeezing parameter $r=0$, thus $G_s=G$ and $\delta_s=\tilde{\delta}$. In presence of optical nonlinearities the situations changes: for simplicity we consider the system to be in the weak optomechanical coupling regime $\tilde{g}_s \ll \kappa,\Gamma $ allowing us to approximate Eqs.~\eqref{eq:OMfreqShift} and \eqref{eq:OMlinewidth} with their value at the unperturbed mechanical frequency $\Omega$. When $\chi > 0$ ($\chi < 0$) the squeezing parameter is negative (positive) at the Stokes sideband $\delta_s=+\Omega$, implying that the scattering process is enhanced $G_s>G$ (suppressed $G_s<G$); for the Anti-Stokes process the scenario is opposite. We also notice that the optical sideband detuning from the cavity mode becomes density-dependent: extremizing $\delta\Gamma_{\Omega}(\delta_s)$ we get the rescaled sideband detunings:
\begin{equation}
	\delta_{s,\pm}=\pm \frac{1}{2\sqrt{3}}\sqrt{4\Omega^2-\kappa^2+2\sqrt{\kappa^4+4\kappa^2\Omega^2+16\Omega^4}},
\end{equation} 
that reduces to $\delta_{s,\pm}\approx\pm\Omega$ in the sideband-resolved regime $\Omega\gg \kappa$. Using the fact that $\chi\approx\tilde{\chi}$ in our system (cf. Appendix G) and inserting Eq.~\eqref{eq:SqueezingParameter} in \eqref{eq:SqueezedFieldDetuning}, one can derive the detuning $\tilde{\delta}_{sb}$ of the two sidebands with respect to the bare oscillator frequency renormalized by the total interaction energy ($\omega_{\tilde{n}}=\omega_0+2\tilde{\chi}\tilde{n}$), yielding 
\begin{equation}\label{eq:SidebandDetuning}
	\tilde{\delta}_{sb}=(\omega_{sb}-\omega_{\tilde{n}})=\pm\sqrt{\delta_{s,\pm}^2+\tilde{\chi}^2\tilde{n}^2}.
\end{equation}
The calculation of the (symmetrized) mechanical displacement spectral density is straightforward in frequency domain. Indeed, provided the dynamics is stable (i.e. $\delta q_{t}$ describes a stationary process), we can use the Wiener-Khinchin theorem to write
\begin{equation}\label{eq:MechdispPSD}
	\overline{S}_{qq}(\omega)=\int \frac{\mathrm{d}t}{2} \, e^{i\omega t} \langle\lbrace\delta q_t , \delta q_{0}\rbrace\rangle = \int \frac{\mathrm{d}\omega'}{4\pi} \langle\lbrace\delta q_{\omega} , \delta q_{\omega'}\rbrace\rangle,
\end{equation}
where $\lbrace \cdot,\cdot\rbrace$ denotes the anti-commutator and we have used the spectral representation of the Dirac delta function $2\pi\delta(t')=\int \mathrm{d}\omega' e^{i\omega' t'}$. Next, one needs inserting Eq.~\eqref{eq:dispSpectrum} in \eqref{eq:MechdispPSD} and Fourier-transform the input noise correlators, finding
\begin{equation}
	\begin{aligned}
		&\langle \xi_m(\omega)\xi_m(\omega') \rangle = 4\pi\tilde{m}\omega\Gamma (\overline{n}_{th}(\omega)+1)\delta(\omega+\omega'),\\
		&\langle \vec{\xi}_s(\omega)\vec{\xi}_s(\omega')^{T}\rangle =2\pi \mathcal{C}_s \delta(\omega+\omega'),
	\end{aligned}	
\end{equation}
where $\overline{n}_{th}(\omega)$ is the thermal occupation of a bosonic mode at a frequency $\omega$ and $\mathcal{C}_s$ is the correlation matrix for the squeezed optical bath given in Eq.~\eqref{eq:CorrMatrix_SqueezedBath}. As the Dirac deltas $\delta(\omega+\omega')$ trivialize the integral in Eq.~\eqref{eq:MechdispPSD}, one finds that
\begin{widetext}
	\begin{equation}\label{eq:DisplPSD}
		\overline{S}_{qq}(\omega)=\vert\Xi_\mathrm{eff}(\omega) \vert^2\left[ \tilde{m}\omega\Gamma \coth\left(\frac{\hbar\omega}{2 k_B T}\right) +\frac{\kappa}{2}\tilde{n}G^2_s[(n_s+\frac{1}{2})(|\Xi_s(\omega)|^2+|\Xi_s(-\omega)|^2)+2 m_s\mathrm{Re}(\Xi_s(\omega)\Xi_s(-\omega))] \right].
	\end{equation}	
\end{widetext}
Following ref.~\cite{Genes2008}, in order to evaluate the efficiency of back-action cooling, we have to find the mean internal energy in the steady-state, which is ultimately related to the effective stationary number of vibrational quanta in the resonator ($\hbar=1$):
\begin{equation}\label{eq:Osc_IntEnergy}
	\begin{aligned}
		\langle H \rangle &= \Omega(n_\mathrm{eff}+1/2) =\int \frac{\mathrm{d}\omega}{2\pi}\left( \frac{\langle p^2_{\omega}\rangle}{2\tilde{m}}  +\frac{1}{2}\tilde{m}\Omega^2\langle q_{\omega}^2\rangle \right)\\
		&=\int \frac{\mathrm{d}\omega}{4\pi} \tilde{m}(\Omega^2 +\omega^2)\overline{S}_{qq}(\omega).
	\end{aligned}	
\end{equation}

Restricting ourselves to the weak optomechanical coupling regime, one can approximate the square modulus of the effective susceptibility with 
\begin{equation}
	\vert\Xi_\mathrm{eff}(\omega) \vert^{2} \approx [\Omega^2\Gamma_{om}^2+(\omega^2-\Omega\,\Omega_{om})^2]^{-1},
\end{equation}
where $\Gamma_{om}=\Gamma+\delta\Gamma_{\Omega}$ and $\Omega_{om}=\Omega+2\delta\Omega_{\Omega}$. The approximate spectral density presents eight poles located at $\omega=\pm \sqrt{\Omega(\Omega_{om}\pm i\Gamma_{om})}$ and at $\omega=\pm \delta_s \pm i\kappa/2$, and a removable singularity at $\omega=0$. We evaluate the integral in the second line of Eq.~\eqref{eq:Osc_IntEnergy} using the residual theorem on the upper half complex plane.
\begin{widetext}
	\noindent The residuals associated to the poles of the optical susceptibility are
	\begin{equation}
		\mathcal{R}_{1,2} = i\frac{g_s^2 n \left(\kappa \pm 2 i \delta _s-2 \Omega \right) \left(\kappa \pm 2 i \delta _s+2 \Omega \right) \left(2 \kappa  m_s+\left(2 n_s+1\right) \left(\kappa \pm 2 i \delta
			_s\right)\right)}{\pi  \left(\kappa \pm 2 i \delta _s\right) \left(16 \Omega ^2 \Gamma _{\text{om}}^2+\left(4 \Omega  \Omega _{\text{om}}+\left(\kappa \pm 2 i \delta
			_s\right){}^2\right){}^2\right)},
	\end{equation}  
	while those associated to the effective mechanical susceptibility read
	\begin{equation}
		\mathcal{R}_{3,4} = \frac{(\Omega^2+\Omega_{\pm}^2)\left( \frac{\Gamma\Omega^2_{\pm}}{\Omega}\coth\left(\frac{\Omega_{\pm}}{2k_B T}\right)A_{\pm} + 8\kappa\tilde{g}_s^2\Omega_{\pm}B_{\pm}\right)}{i 16\pi\Omega\,\Omega_{\pm}^2\Gamma_{\mathrm{om}} A_{\pm}},
	\end{equation}
	with
	\begin{equation}
		\begin{aligned}
			\Omega_{\pm} &=\pm\sqrt{\Omega(\Omega_{\mathrm{om}}\pm\Gamma_{\mathrm{om}})},\\
			A_{\pm}& =(4\Omega^2_{\pm}+(\kappa-2i\delta_s)^2)(4\Omega^2_{\pm}+(\kappa+2i\delta_s)^2),\\
			B_{\pm}& =(2m_s(\kappa^2-4\delta_s^2+4\Omega_{\pm}^2)+2 n_s + 1)(\kappa^2+4\delta_s^2+4\Omega_{\pm}^2)).
		\end{aligned}
	\end{equation}
\end{widetext}

We can thus express the internal energy as $\langle H\rangle=-2\pi i \sum_{j} \mathcal{R}_j$, and deduce the analytical expression for the effective mean occupation of the resonator in the steady-state using $n_\mathrm{eff}=(\langle H\rangle/\Omega-1/2)$. 	The result of the above calculation used to trace the solid lines in Fig.~\ref{fig:NonlinearSBcool}\,(c,f) of the main text.

In Fig.~\ref{fig:NonlinearSBcool}\,(d) we trace the minimum average phonon occupation (i.e. at the optimal laser detuning $\tilde{\delta}_{sb}^{-}$) as a function of cavity occupation. For a frequency-softening Kerr interaction ($\chi<0$), we show that a finite yet modest cooling enhancement is possible when compared to the harmonic-resonator case ($\chi=0$). A natural question is whether this enhancement is parameter-dependent: based on a qualitative scaling argument, in the main text, we indicate that in the bad cavity limit $\kappa_l/\Omega\gg 1$ the effect may become sizable.

In the following, we want to quantitatively address this matter. In order to ease a bit the algebra, we restrict ourselves to the case where a perturbative treatment based on Fermi's golden rule is applicable~\cite{Marquardt2007}. From the right-hand side of Eq.~\eqref{eq:DisplPSD}, it is straightforward to extract the spectrum of the radiation-pressure force $S_{FF}(\omega)$. The net optical damping is then $\Gamma_{\mathrm{opt}}=x^2_{\mathrm{ZPF}} [S_{FF}(+\Omega)-S_{FF}(-\Omega)]$, i.e. Eq.~\eqref{eq:OMlinewidth} for $\omega=\Omega$. The effective steady-state phonon occupation is then given by~\cite{Aspelmeyer2014,Marquardt2007}
\begin{figure}[tb]
	\centering
	\includegraphics[trim=0cm 0cm 0cm 0cm, width=70mm]{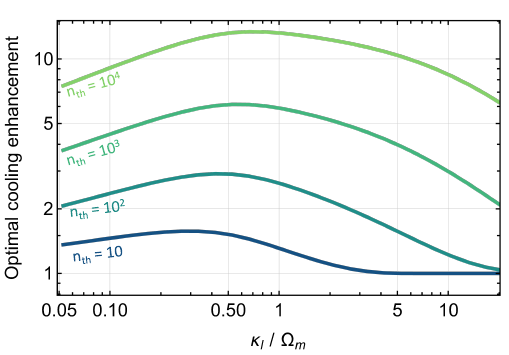}
	\caption{Nonlinear cooling enhancement: Dependence of $\eta_{cool}$ on the polariton linewidth and on the thermal phonon occupation ($n_{\mathrm{th}}$). The curves are obtained by numerical minimization of $\eta_{cool}^{-1}$ as a function of $\tilde{n}$.}
	\label{fig:CoolingENH}
\end{figure}

\begin{equation}
 n_{\mathrm{eff}}=\frac{x^2_{\mathrm{ZPF}}S_{FF}(-\Omega)+\Gamma n_{\mathrm{th}}}{\Gamma_{\mathrm{opt}}+\Gamma}, 
\end{equation}
where $n_{\mathrm{th}}$ denotes the number of excitation at thermal equilibrium in the absence of sideband cooling. Substituting in the above expression the squeezed-oscillator parameters $(\delta_s,n_s,m_s)$, assuming $\tilde{\chi}\approx\chi$, using the identity $\delta_s m_s=-\chi\tilde{n}/2$, and performing some algebra with the hyperbolic functions, we obtain the following result
\begin{equation}\label{eq:OptCoolingFGR}
     n_{\mathrm{eff}}=\frac{\tilde{C}_{lm} F_1(\tilde{\delta}) +  n_{\mathrm{th}}\kappa^{-2} F_2(\tilde{\delta})}{ \tilde{C}_{lm} F_3(\tilde{\delta})+ \kappa^{-2}F_2(\tilde{\delta})},
\end{equation}
where $\tilde{C}_{lm}=C_{0} \tilde{n}=4g_{lm}^2\tilde{n}/\kappa\Gamma$ and the short-hand notations $F_{i}(\tilde{\delta})$ stand for
\begin{equation}
\begin{aligned}
    F_1&=4(\tilde{\delta}^2+\chi\tilde{n}+\Omega)^2+\kappa^2\frac{\tilde{\delta}+\chi\tilde{n}}{\tilde{\delta}-\chi\tilde{n}},\\
    F_2&=16(\tilde{\delta}^2-\chi^2\tilde{n}^2-\Omega^2)+8\kappa^2(\tilde{\delta}^2-\chi^2\tilde{n}^2+\Omega^2)+\kappa^4,\\
    F_3&=16 (\tilde{\delta}+\chi\tilde{n})\Omega.
\end{aligned}    
\end{equation}
These relations can be used to define the cooling enhancement for any given set of parameters. In order to evaluate the optimal cooling enhancement, we substitute in Eq.~\eqref{eq:OptCoolingFGR} the expression for the optimal laser detuning given by Eq.~\eqref{eq:SidebandDetuning}. The resulting expression is cumbersome but analytic. Finally, one can take the ratio between the optimal phonon occupations in the linear ($\chi=0$) and nonlinear case ($\chi\neq 0$) and define the optimal cooling enhancement as $\eta_{\mathrm{cool}}=\mathrm{max}_{\tilde{n}}|n^{\mathrm{opt}}_{\mathrm{eff}}(0,\tilde{n})/n^\mathrm{opt}_{\mathrm{eff}}(\chi,\tilde{n})|$. The function has always a maximum larger than unity in the $\chi<0$ case. Figure~\ref{fig:CoolingENH} shows the optimal nonlinear sideband cooling enhancement ($\eta_{\mathrm{cool}}$) as a function of the polariton-mode linewidth and thermal-phonon occupation. In the bad-cavity limit, and for large numbers of thermal phonons; an order of magnitude enhancement in the cooling performance is predicted, in good agreement with recent experiments performed on a nonlinear cavity-magnetomechanical device~\cite{Zoepfl2022}.

%%%%

\subsection{K - Phonoritons\label{app:K}}	

For a large intracavity population $\tilde{n}$ the effective optomechanical 	coupling strength $g_s^2\tilde{n}$ eventually exceeds the both polariton decay rate $\kappa$ and the mechanical damping rate $\Gamma$. The coherent exchange of energy between the two degrees of freedom becomes so fast that the normal modes of the system hybridize. As a consequence the fundamental excitations of the system are weighted superposition of light, matter and sound, which are sometimes referred as phonoritons. As the optomechanical gain exceeds unity in vicinity of the anti-stokes sideband (cf.~Fig.~\ref{fig:NonlinearSBcool}\,(a) in the main text) yielding mechanical self-oscillations that are not correctly described by our linearization hypothesis, we will concentrate on the effects of strong coupling at the Stokes sideband.

\begin{figure}[tb]
	\centering
	\includegraphics[trim=0cm 0cm 0cm 0cm, width=86mm]{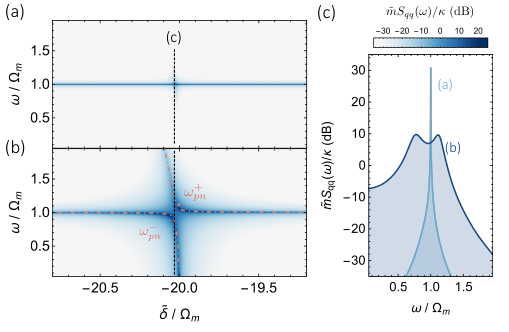}
	\caption{Phonoritons: Displacement power spectral density as a function of the rescaled detuning $\tilde{\delta}$ obtained for a large intracavity population $\tilde{n}=\num{2.0e3}$, in the case of a repulsive (a) or an attractive (b) optical nonlinearity. All parameters are chosen as in Fig.~\ref{fig:NonlinearSBcool} in the main text. In the case of attractive nonlinearities the effective optomechanical coupling exceeds the polariton decay rate $\kappa$, yielding the characteristic anti-crossing of the normal modes $\omega_{pn}^{\pm}$ about $\omega=\Omega_m$. (c) Representative spectral densities for a detuning $\tilde{\delta}/\Omega_m=-20.02$.}
	\label{fig:Phonoritons}
\end{figure}

In Fig.~\ref{fig:Phonoritons}\,(a) we plot the displacement power spectral density [Eq.~\eqref{eq:DisplPSD}] as a function of the rescaled detuning $\tilde{\delta}$ for a intracavity population $\tilde{n}=2000$ an the other parameters as in the main text ($\kappa/2\pi=\SI{6.5}{\GHz}$, $\Omega_m/\kappa=3$, $\Gamma/\kappa=\num{1e-4}$, $g/\kappa=\num{2e-3}$, $\chi/\kappa=\num{3e-2}$). As the effective laser drive detuning crosses the Stokes resonance for $\tilde{\delta}= -\tilde{\delta}_{sb}\approx -20 \Omega_m$ we see a crossing of spectral features: the system is in the weak coupling regime. Instead, if now we plot the same density map but for attractive nonlinearities ($\chi/\kappa=\num{-3e-2}$) one can observe an avoided crossing that is the signature of the strong coupling regime. In Fig.~\ref{fig:Phonoritons}\,(c) we show the profile of the displacement power spectrum at the Stokes resonance, indicated by the dashed black line in Fig.~\ref{fig:Phonoritons}\,(a,b). The fact that depending on the sign of the nonlinearity the system can either be in the weak or strong coupling is not surprising as $g_s$ explicitly depends on $\chi$, as shown in Fig.~\ref{fig:NonlinearSBcool}\,(b) of the main text.

In order to describe the frequency of the normal modes $\omega_{pn}^{\pm}$, we again rely on the squeezing transformation to map the known results for the case of an harmonic cavity mode~\cite{Aspelmeyer2014} to the case of Kerr nonlinearities. Neglecting the mechanical damping $\Gamma/\kappa\ll 1$ one has
\begin{equation}
	\omega_{pn}^{\pm}=\frac{(\Omega_m-\delta_s)}{2}\pm\sqrt{\tilde{n}g^2_s+\frac{(\Omega_m+\delta_s+i\kappa/2)^2}{4}},
\end{equation}
which we trace for comparison as a dashed line in Fig.~\ref{fig:Phonoritons}\,(b). Using Eq.~\eqref{eq:SidebandDetuning}, we can write the normal mode splitting at the Stokes sideband in the sideband resolved limit as
\begin{equation}
	\Delta\omega_{pn}=2\sqrt{\tilde{n}g^2\eta_{-}-\kappa^2/16},
\end{equation}
where we recall that $\eta_{-}=(1+\chi\tilde{n}/\tilde{\delta}_{-})^{1/2}/(1+\chi\tilde{n}/\tilde{\delta}_{+})^{1/2}$ is the sideband enhancement factor introduced in the main text.

\end{document}